\documentclass[usenatbib]{rspublic}
\usepackage{graphicx}
\usepackage{bm}
\usepackage{color}

\renewcommand{\phi}{\varphi}
\renewcommand{\epsilon}{\varepsilon}

\begin{document}
\title[Hard discs under steady shear]{Hard discs under steady shear: comparison of Brownian dynamics simulations and Mode Coupling Theory}

\author[O. Henrich, F. Weysser, M. E. Cates, M. Fuchs]{Oliver Henrich$^{1}$, Fabian Weysser$^{2}$, Michael E. Cates$^{1}$, Matthias Fuchs$^{2}$}

\affiliation{$^{1}$School of Physics and Astronomy, University of
Edinburgh, JCMB, The King's Buildings, Edinburgh EH9 3JZ, UK \\
$^{2}$Universit\"at Konstanz, Fachbereich Physik, D-78457 Konstanz,
Germany}

\label{firstpage} \maketitle

\begin{abstract}{}
Brownian dynamics simulations of bidisperse hard discs moving in two dimensions in a given steady and homogeneous shear flow are presented close to and above the glass transition density. The stationary structure functions and stresses of  shear melted glass are compared quantitatively to parameter free numerical calculations for monodisperse hard discs using mode coupling theory (MCT) within the integration through transients (ITT) framework. Theory qualitatively explains the properties of the yielding glass but quantitatively overestimates the shear driven stresses and structural anisotropies.
\end{abstract}

\section{Introduction}

Shear flow can drive dense colloidal dispersions into states far
from equilibrium. Especially of interest is the possibility to shear
melt colloidal solids, in particular metastable colloidal glasses
and gels, and to investigate  shear-melted (yielding) colloidal
glasses. Does a yield stress and/or yield strain exist (Petekidis {\it et al.} 2002)?
Are shear-melted states necessarily heterogeneous (e.g. shear
banded)? Does ageing prevent stationary states under steady
shearing? Do hydrodynamic interactions dominate the properties in
flow? These are just some of the questions whose answers will
provide insights into the still murky glassy state.

Recently the integration through transients (ITT) approach has been
used to generalize the mode coupling theory (MCT) of the structural
glass transition to the case of steady shearing
(Fuchs \and Cates 2002, 2009), and to arbitrary homogeneous flows
(Brader {\it et al.} 2008), albeit based on a number of approximations. First,
homogeneous flow profiles are assumed from the outset, second,
hydrodynamic interactions are neglected, and third, the mode
coupling (mean-field-like) decoupling approximation, splitting a
four point density fluctuation function into the product of
two-point density correlators, has been applied. While simplified
(schematic) models of MCT-ITT (Fuchs \and Cates 2003)
can well be fitted to the linear rheology
(frequency dependent storage and loss moduli) and non-linear
rheology (flow curves, viz. stress as function of shear rate) of
model dispersions (Siebenb\"urger {\it et al.} 2009), no fully quantitative comparison between 
the theory and data without additional approximations has been
presented up to now.

We provide this first quantitative comparison, solving the MCT-ITT
equations for hard discs confined to move in a plane, and performing
Brownian dynamics simulations for a binary hard disc mixture, also
in two dimensions. Computational (memory) constraints force us to work in two dimensions, 
which is still of some experimental relevance, as glasses have been observed in 
quasi-two-dimensional dispersions (Bayer {\it et al.} 2007). Use of a binary mixture suppresses the
nucleation rate sufficiently for even our long simulation runs to
remain free of crystal nuclei. As hydrodynamic interactions are
absent also in the simulation, and as Lees-Edwards boundary
conditions (LE) in combination with the thermostat impose a homogeneous shear rate, 
the integration of the equations of motion in the simulation ensure a stringent test of the theory.
Results for the stationary shear and normal stresses, and for the
shear-distorted microstructure are presented and discussed. The
input quantities entering the theory, and transient density
correlation functions, which (crucially) enter during intermediate
MCT-ITT steps, are characterized also.

Our work bears some similarity to the study by Miyazaki, Reichman,
and Yamamoto, who, however, concentrated on  fluid states under
shear, and on their time dependent fluctuations (Miyazaki {\it et al.} 2004). We
focus here on the yielding glass state, and its stationary,
time-independent averages, which are not accessible to the theory in
Ref. (Miyazaki {\it et al.} 2004). In order to compare quantitatively with theory,
we aimed for better statistics than in comparable previous two-dimensional simulation
studies of sheared glasses (Yamamoto \and Onuki 1998; Furukawa {\it et al.} 2009).

\section{Simulation}

The basic concept of the algorithm has been described in detail in
three dimensions in (Scala {\it et al.} 2007) and can easily be adapted to two
dimensions $x$ and $y$. In order to prevent crystallisation at high densities 
we consider a binary mixture with a diameter ratio of $D_2/D_1=1.4$ at equal 
number concentrations ($N_{\rm small}=N_{\rm big}$). $N=1000$ hard discs move in a 
simulation box of volume $V$ with periodic Lees Edwards boundary conditions at 
packing fraction $\varphi = \frac{N \pi}{8 V} \cdot (D_1^2 + D_2^2)$. 
The mass of the particles is set equal to unity, thermal energy  $k_BT$ sets the energy scale, and $D_1\equiv D$ is used as unit of length in the following. We choose our coordinate axes such
that flow is in the $x$-direction and the shear gradient $\dot \gamma$ is in the
$y$-direction.  
After putting the particles on their initial positions we provide Gaussian distributed velocities which will be overlain by the linear shear flow.
To propagate the system forwards in
time we employ a semi-event-driven algorithm. For every particle at
the time $t_0$ the algorithm determines the possible collision time
$\tau$ with any other particle. This is easily achieved by solving the
equation
\begin{equation}
\frac{1}{2}(D_i + D_j) = |\bm{r}_{ij}- {\bm v}_{ij} \; \tau|
\end{equation}
where $D_i$ and $D_j$ denote the diameters of the particles $i$ and $j$, ${\bm
r}_{ij}$ the relative vector between both particles, and ${\bm
v}_{ij}$ the relative velocities. The smallest solution $\tau_{min}$
for all particle pairs determines the next event in the algorithm. All
particles can then be propagated according to: ${\bm r}(\tau_{min})= {\bm r}(t_0)+{\bm v} \; \tau_{min}$.
With the conservation of energy and momentum the binary collision
laws impose new velocities ${\bm v}_j^*$ for the particle $j$ and
$i$
\begin{eqnarray}
{\bm v}_j^* = {\bm v}_j + (\hat{\bm{r}}_{ij} \cdot {\bm
v}_{ij})\hat{\bm{r}}_{ij}\; , \qquad {\bm v}_j^* = {\bm v}_j -
(\hat{\bm{r}}_{ij} \cdot {\bm v}_{ij})\hat{\bm{r}}_{ij}.
\end{eqnarray}
Due to the boundary conditions any particle in the vicinity of the
box-boundary can collide with an image particle coming from the
other end of the box. The boxes are constructed in such a way that
they are translated with the velocity $\dot{\gamma}L$, where $L$
denotes the size of the box. The boundary conditions thus have the
consequence that the velocities tend to
increase. Therefore a thermostat has to be introduced. After a time
$\tau_B = 0.01$ a so-called Brownian step sets in, which assures
that the particles move diffusively for longer times. In the
Brownian step at the time $\tau_B$, all particle velocities are
freshly drawn from a Gaussian distribution with $\langle |{\bm
v}|^2\rangle=2$. After that linear shear flow is imposed so that
$\langle {\bm v} \rangle=\dot{\gamma} y(\tau_B) \hat{\bf x}$ holds.

As the system starts from a cubic lattice it is necessary to wait
for the system to relax before meaningful stationary averages can be
taken. The quantity of interest in the work presented here is the
potential part of the stress: $ \sigma_{\alpha \beta}(\dot\gamma) =
\frac{1}{V} \langle \sum_{ij} ({\bm F}(t)_{ij})_\alpha ({\bm
r}(t)_{ij})_\beta \rangle^{(\dot\gamma)}$, with  the relative
force components (in $\alpha$-direction) of particles $i$ and $j$
$({\bm F}(t)_{ij})_\alpha$, and the particles' relative distance
components $({\bm r}(t)_{ij})_\beta $. The kinetic part will play
no role, and thus has already been omitted. As we consider hard
particles the forces must be calculated from the collision events.
By observing the collisions within a certain time window $\Delta
\tau_c$ forces may be extracted using the change of momentum which
occurs during the observation time. This leads to the evaluation
algorithm (Lange {\it et al.} 2009; Foss \and Brady 2000)
\begin{equation}\label{coll_sigma}
 \sigma_{\alpha \beta}(\dot\gamma) = \left \langle  \frac{1}{\Delta \tau_c}
 \sum_{t_c \in \lbrace t, t+\Delta \tau_c \rbrace}\;  (\Delta {\bm v}(t_c)_{ij})_\alpha ({\bm r}(t_c)_{ij})_\beta \right \rangle_s\; ,\\
\end{equation}
where the summation is over all collisions of particles $i$ and $j$ at time $t_c$ within the time
window $\Delta \tau_c$. The procedure effectively sums the momentum
changes $(\Delta {\bm v}_{ij})_\alpha$ in the $\alpha-$ direction
multiplied by the relative distance of the particles $({\bm
r}_{ij})_\beta$ in the $\beta-$ direction.
Here and below the brackets $\langle...\rangle_s$ denote the  average
over different simulation runs.\\
\noindent Additionally the shear stress can be computed via the contact value $g({\bm r})$:
\begin{equation}
\label{contactvalue}
\sigma_{xy}(\dot \gamma) = \frac{8 k_B T}{\pi^2 } \phi^2 \sum \limits_{i,j \in \lbrace 1,2 \rbrace} \frac{1}{d_{ij}^2} \int {\rm d} \theta \cos(\theta) \sin(\theta) g^{ij}(r,\theta) 
\end{equation}
where $g^{11}(r, \theta), g^{12}(r, \theta)...$ denote the $\theta$-dependent partial contact values of the two components and $d_{11}, d_{12},...$ the minimal distance between two particles. $\theta$ is the polar angle.\\
The Green-Kubo relation $\eta_{xy} = \frac{1}{k_B T V} \int \limits_{0}^{\infty} \langle \sigma_{xy}(t)  \sigma_{xy}(t+s) \rangle {\rm d}s $ holds for the non-sheared system. Thus the shear viscosity can be extracted from the simulation via (Alder {\it et al.} 1970)
\begin{equation} \label{shearviscosity}
\eta_{xy} = \frac{1}{2 k_B T V} \lim \limits_{t \to \infty} \frac{1}{t} \left \langle \left(  \int \limits_{0}^{\infty} \sigma_{xy}(t') {\rm d}t' \right)^2 \right \rangle = \frac{1}{V} \left \langle \left( \sum \limits_{coll} {\bm r}_y \Delta {\bm v}_x \right)^2 \right \rangle_s .
\end{equation}
where the sum runs over all collisions.
The second pivot in this
analysis is the equal-time structure factor and its deviation from the
quiescent system. Exploiting that $S_{\bm q}(\dot\gamma) = 
\langle \frac{1}{N} \sum_{i,j} \exp(i {\bm q}({\bm r}_i -
{\bm r}_j))  \rangle^{(\dot\gamma)}$ is a real quantity, we
can extract it via
\begin{equation}
S_{\bm q}(\dot\gamma) =  \left \langle \frac{1}{N} \sum
\limits_{(i,j) \in NI} \cos({\bm q}({\bm r}_i - {\bm r}_j)) \right
\rangle_s
\end{equation}
where the double sum runs over all pairs of particles $i$ and $j$ 
($NI$). The pairs $ij$ are determined by the LE boundary conditions and 
the constraint of having the lowest distance among all possible image particles 
in the surrounding boxes.\\
For low Peclet numbers (Pe$_0$=$\dot\gamma D^2/D_0$)in fluid states, $g({\bm r})$ can be expanded: $g({\bm r}) = g_0(r) + 2 {\rm Pe}_0 \frac{xy}{r^2} g_1(r) + \mathcal{O} ({\rm Pe}_0^2)$. This result can be used to derive the relative distortion of the structure factor in the linear response regime (Strating 1999).
\begin{equation}
\delta S({\bm q}, \dot{\gamma} \to 0) = 2 {\rm Pe}_0 \int {\rm d}r \;  r g_1(r) \int {\rm d} \theta \; \cos(\theta) \sin(\theta) {\rm e}^ {[i (r q_x \cos (\theta) + r q_y \sin (\theta))]}
\end{equation}
while $g_1(r)$ can be obtained via
\begin{equation}
\frac{ 2 \pi}{{\rm Pe}_0} \int {\rm d} \theta \;  g({\bm r}) \cos(\theta) \sin(\theta) = g_1(r) + \mathcal{O}(¸{\rm Pe}_0^2 \cdot xy/r^2).
\end{equation}
\section{Mode Coupling Theory in the integrations through transients framework}

The MCT-ITT approach generalizes the MCT of the glass transition to
colloidal dispersions under strong continuous shear. It considers
$N$ spherical particles with arbitrary interaction potential which
move by Brownian motion relative to a given linear shear profile.
An equation of motion for a {\em transient} density correlator
$\Phi_{\bf q}(t)$ encodes structural rearrangements, and
approximated generalized Green-Kubo laws relate stress relaxation to the
decay of density fluctuations.

The transient density correlator is defined by $\Phi_{\bf
q}(t)=\langle \rho_{\bf q}^* \rho_{{\bf q}(t)}(t) \rangle / N S_q$,
where the density fluctuation is as usual $\rho_{\bf q}(t) =
\sum_{j=1}^N\, \exp{\{ i {\bf q}\cdot {\bf r}_j(t)\}} $, and, in
ITT, the average can be performed over the equilibrium
Gibbs-Boltzmann ensemble. Thus the normalization of the initial
value $\Phi_{\bf q}(0)=1$ is given by the equilibrium structure
factor $S_q$.  The time-dependent or {\em shear-advected} wavevector
${\bf q}(t)= (q_x,q_y-\dot\gamma t q_x)^T$ appearing in the
definition eliminates the  affine particle motion with the flow
field, and gives $\Phi_{\bf q}(t)\equiv1$ in the absence of Brownian
motion. Shear flow coupled to random motion causes $\Phi_{\bf q}(t)$
to decay, as given by an (exact) equation of motion containing a
retarded friction kernel which arises from the competition of
particle caging and shear advection of fluctuations
\begin{equation}\label{mct1}
\dot{\Phi}_{\bf q}(t) + \Gamma_{\bf q}(t) \; \left\{ \Phi_{\bf q}(t)
+ \int_0^t dt'\; m_{\bf q}(t,t') \; \dot{\Phi}_{\bf q}(t') \right\}
=0 \;,
\end{equation}
where the initial decay rate contains Taylor dispersion as $\Gamma_{\bf
q}(t) = q^2(t)/ S_{q(t)}$. The generalized friction kernel $m_{{\bf
q}}(t,t')$, which is an autocorrelation function of fluctuating
stresses, is approximated, following MCT precepts by an expression involving the structural
rearrangements captured in the density correlators
\begin{eqnarray}\label{mct2} 
m_{\bf q}(t,t') =  \int\!\!\! \frac{d^2k}{(2\pi)^2} \frac{n S_{q(t)}
S_{k(t')}\, S_{p(t')}}{2 q^2(t)\; q^2(t')}\; V_{\bf q k p}(t)\,
V_{\bf q k p}(t')  \; \Phi_{{\bf k}(t')}(t-t')\,  \Phi_{{\bf
p}(t')}(t-t')
\end{eqnarray}
with abbreviation ${\bf p}={\bf q}-{\bf k}$, $n$ the particle density,  and a vertex function given by
\begin{equation}\label{mct3}
V_{\bf q k p}(t)=  {\bf q}(t)\cdot \left( {\bf k}(t) \, c_{k(t)} +
{\bf p}(t)\, c_{p(t)} \right)
\end{equation}
where $c_k$ is the Ornstein-Zernicke direct correlation function
$c_k=(1-1/S_k)/n$. An additional memory kernel is neglected
(Fuchs \and Cates 2009). The equilibrium structure factor, $S_k$, encodes the
particle interactions and introduces the experimental control
parameters like density and temperature. Similarly, the potential
part of the stress $\sigma_{\alpha\beta}(\dot\gamma)=\langle \sigma_{\alpha\beta}
\rangle^{(\dot\gamma)}/V$  in the non-equilibrium stationary state
(neglecting the diagonal contribution that gives the pressure) is
approximated assuming that stress relaxations can be computed from
integrating the transient density correlations
\begin{equation}\label{sigma_xy}
\sigma_{\alpha\beta}(\dot{\gamma})= \frac{n k_B T}{2} \int_0^\infty\!\!\!
dt \int\!\!\! \frac{d^2q}{(2\pi)^2}  \frac{\partial S_{q(-t)}}{\partial t} \frac{q_\alpha q_\beta}{q}
\; \frac{\partial c_q}{\partial q}\; \Phi_{{\bf q}(-t)}^2(t)\;.
\end{equation}
Flow also leads to the build up of 
shear-induced micro-structural changes, which, again integrating up the
transient density correlators, can be found from
\begin{eqnarray}\label{distsq}
S_{\bf q}(\dot{\gamma})&=& S_q +\int_0^\infty dt\; \frac{\partial
S_{ q(-t)}}{\partial t}\; \Phi^2_{{\bf q}(-t)}(t)\; .
\end{eqnarray}
A far smaller isotropic term in $S_{\bf q}(\dot{\gamma})$ (see
(Fuchs \and Cates 2009)) is neglected here, as it is of importance for the
plane perpendicular to the flow only.

\section{Numerical aspects}

The set of coupled Eqs. \ref{mct1},\ref{mct2} and \ref{mct3} was solved self-consistently using
modifications of standard algorithms and the Ng-iteration-scheme
(Ng  2009). The functions were discretized on a 2D-Fourier-grid
consisting of 101 grid points in either coordinate direction using a
cutoff of $qD=30$. More explicitly the Fourier-grid was
discretized as $ q_{x,y}D = -30, -29.4,\dots ,0, \dots,  29.4, 30$
with $D$ being the particle diameter. In order to enhance the
accuracy, the advected direct correlation function $c_{q(t)}$ in Eq.
\ref{mct3} was calculated from the input structure factor once it has
been advected beyond the Fourier-grid cutoff up to $qD=100$. For the
time-integration of the convolution integral in Eq. \ref{mct1} an
initial time step of $10^{-9} D^2/D_0$ was used. The
time-integration in the calculation of derived quantities Eqs.
\ref{sigma_xy} and \ref{distsq}  was performed by dividing the 
integration interval into two sub-intervals, the first one containing times 
shorter, the other one times longer
than $t^\ast=1/(1000 \dot{\gamma})\, D^2/D_0$. 
The integration was then
carried out on the intrinsic quasi-logarithmic grid of the MCT equation solver 
for times $0\le t \le t^\ast$ and on an equally-spaced linear 
time-grid with spacing $\Delta t=t^\ast$ for times $t>t^\ast$.

\section{Equilibrium and transient quantities}

The MCT-ITT approach uses structural equilibrium correlations as
input, computes transient structural density correlators to encode
the competition between flow-induced and Brownian motion, and calculates all
stationary properties from time-integrals over the transient
fluctuations. In this section, the input and intermediate quantities
of the theoretical calculations are presented and discussed.

\subsection{Equilibrium structure factor}

\begin{figure}[htp]
\includegraphics[clip = true, width=0.48\textwidth]{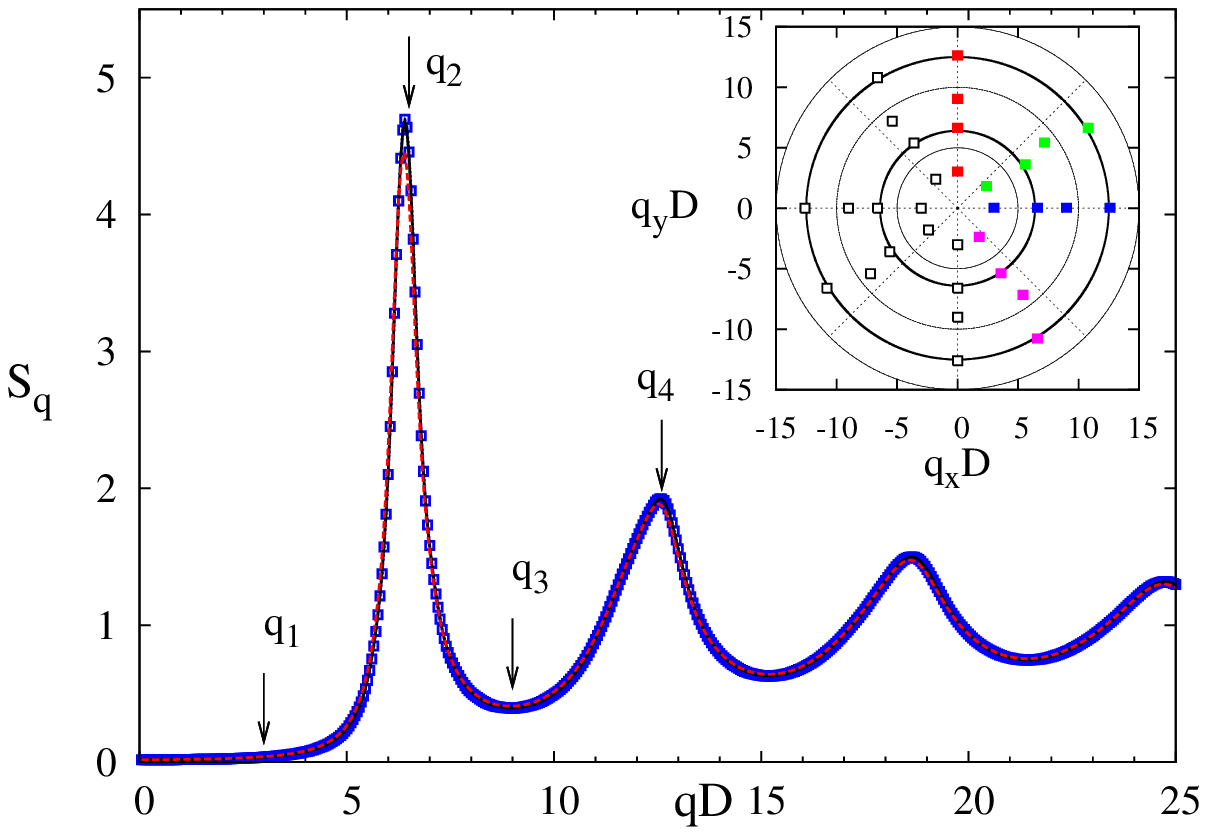}
\includegraphics[clip = true, width=0.48\textwidth]{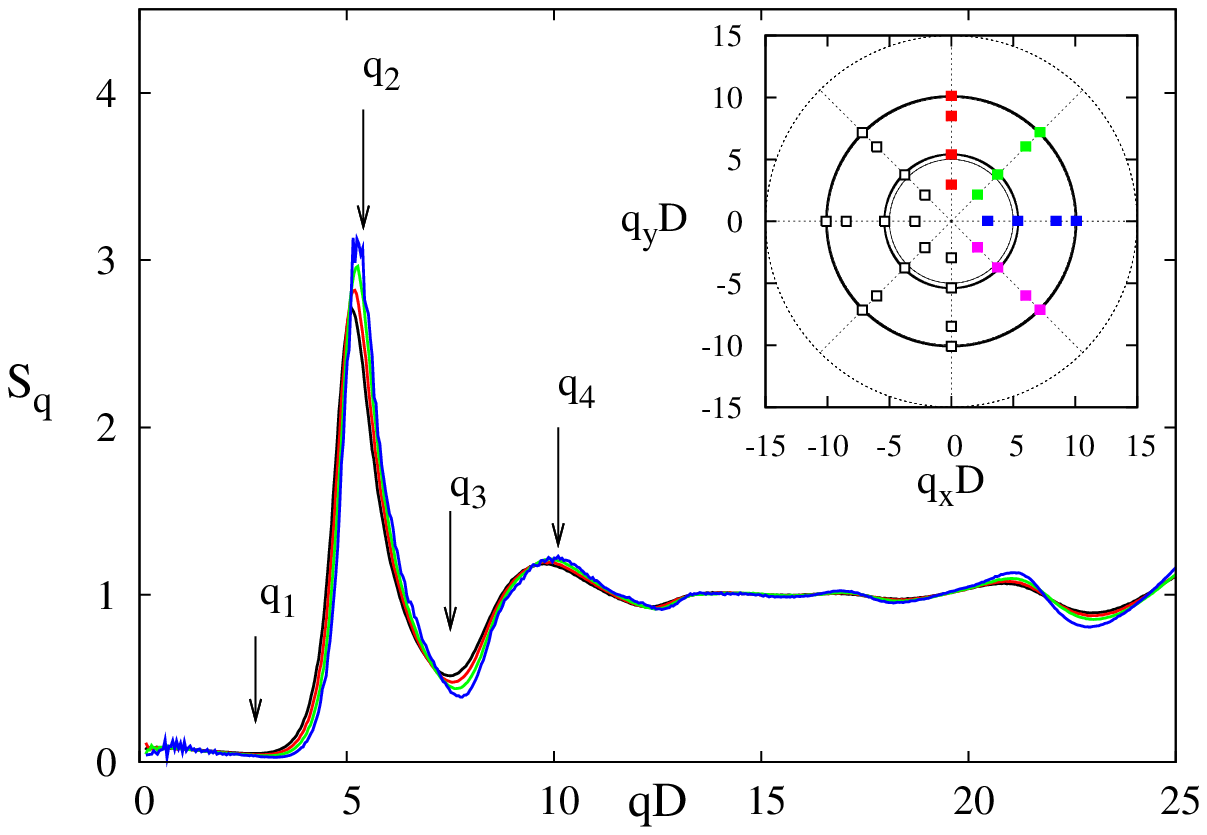}
\caption{Left panel: modified hyper-netted chain structure factor,
used as input quantity to MCT,  for separation parameters
$\epsilon=-10^{-2}$(red), $10^{-5}$ (black) and $10^{-3}$ (blue).
The critical packing fraction is $\varphi_c=0.69899$. The selected
wavevectors are $q_1 D = 3.0$, $q_2 D = 6.6$, $q_3 D = 9.0$ and $q_4
D = 12.6$. Right panel: averaged structure factors for the simulated
binary system. The packing fractions are $\varphi = 0.72$, $0.74$, $0.76$ and $\varphi = 0.79$. The approximate critical
packing fraction is $\varphi_c\simeq 0.79\pm0.005$, as estimated
from the flow curves of Fig. \ref{figstress}. Both insets exhibit
the magnitude and the relative orientation of the selected wavevectors
and introduce the colour code, which is used later on throughout the
comparison. The wavevectors for the right panel are: $q_1 D = 2.8$,
$q_2 D = 5.4$, $q_3 D = 7.5$ and $q_4 D = 10.1$. } \label{fig1}
\end{figure}

The equilibrium structure factor $S_q$ is the only input quantity to
the MCT-ITT equations. It varies smoothly with density or
temperature, but leads to the transition from a shear-thinning fluid
to a yielding glassy state at a glass transition density
$\phi_c$. Figure \ref{fig1} shows modified hyper-netted chain
structure factors of monodisperse hard discs in $d=2$ from
(Bayer {\it et al.} 2007) used in the MCT-ITT calculations. For  comparison, the
averaged structure factors obtained from the binary mixture
simulation are shown also. In both cases, the value of the glass
transition density $\phi_c$ is included in the curves in  Fig.
\ref{fig1}, and only smooth changes in $S_q$ are noticeable. The
short range order at the average particle distance, as measured in
the primary peak of $S_q$, increases with densification. Differences
in the structure between the mono- and the bidisperse system become
appreciable beyond the primary peak, and especially beyond the second
peak in $S_q$. Also the height of the primary peak in $S_q$ at
$\phi_c$ differs, indicating that the averaged structure factor in
the bidisperse system does not well characterize the local structure
and caging in the simpler system, and that a comparison with a
bidisperse MCT-ITT calculation should be performed. As this is
outside the present numerical reach, we proceed by comparing results
for the characteristic wavevectors denoted $q_1$ to $q_4$ in Fig.
\ref{fig1}. The angular dependence of the anisotropic structure will
be explored along the special directions indicated in the insets.

\subsection{Transient coherent density correlation functions}

\begin{figure}[htp]
 \includegraphics[clip = true, width=0.48\textwidth]{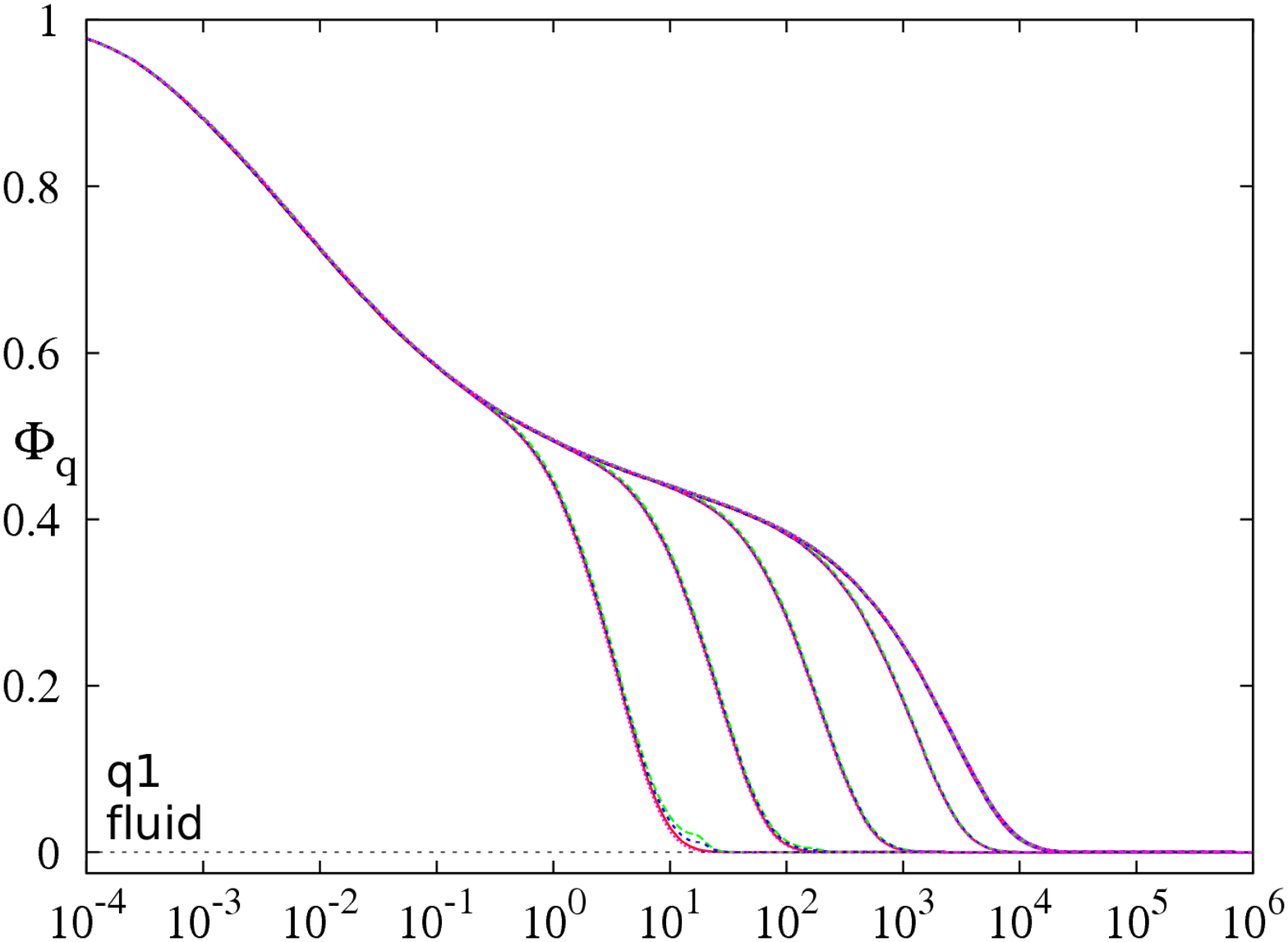}\hfill
 \includegraphics[clip = true, width=0.48\textwidth]{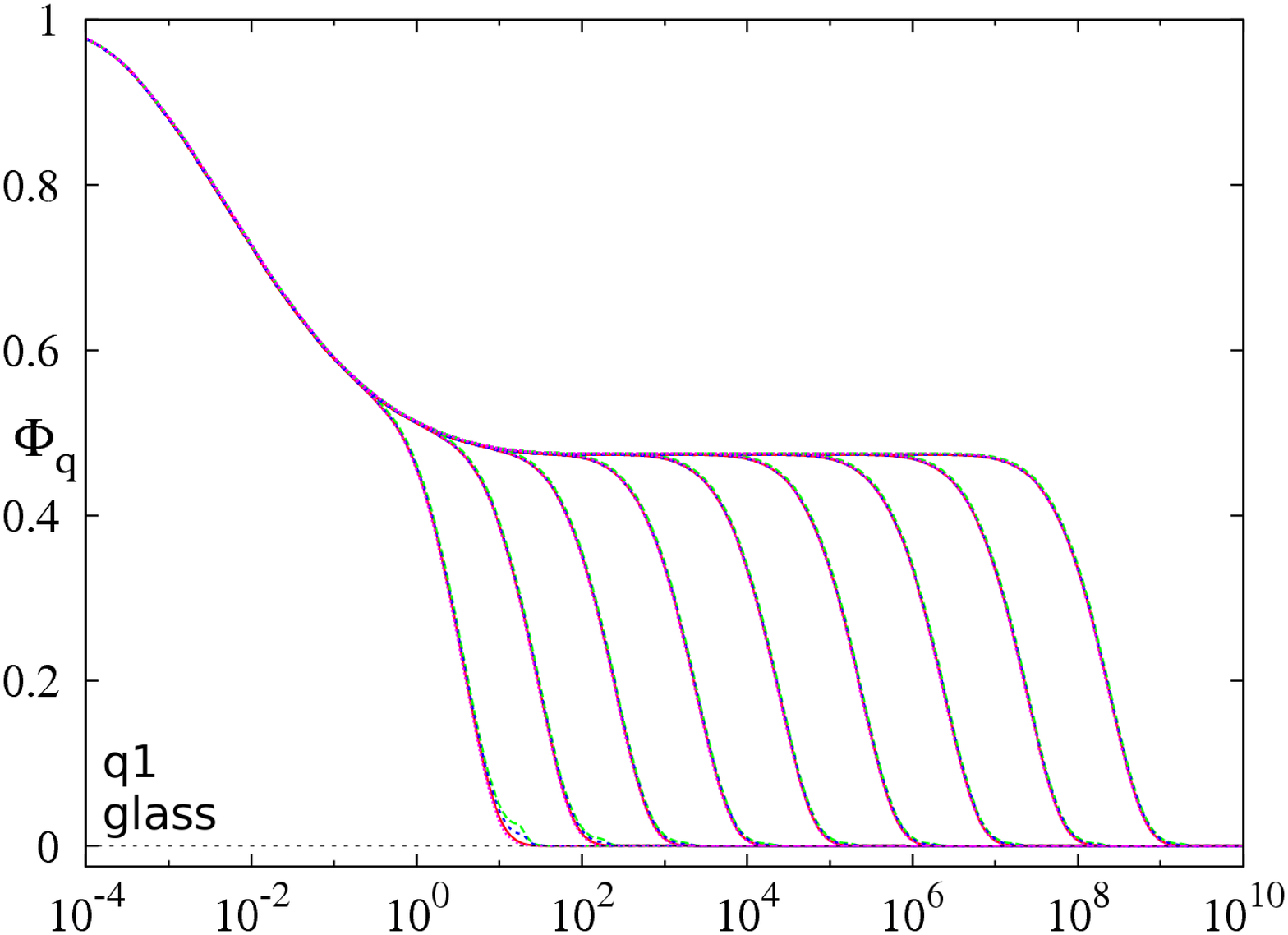}\\
 \includegraphics[clip = true, width=0.48\textwidth]{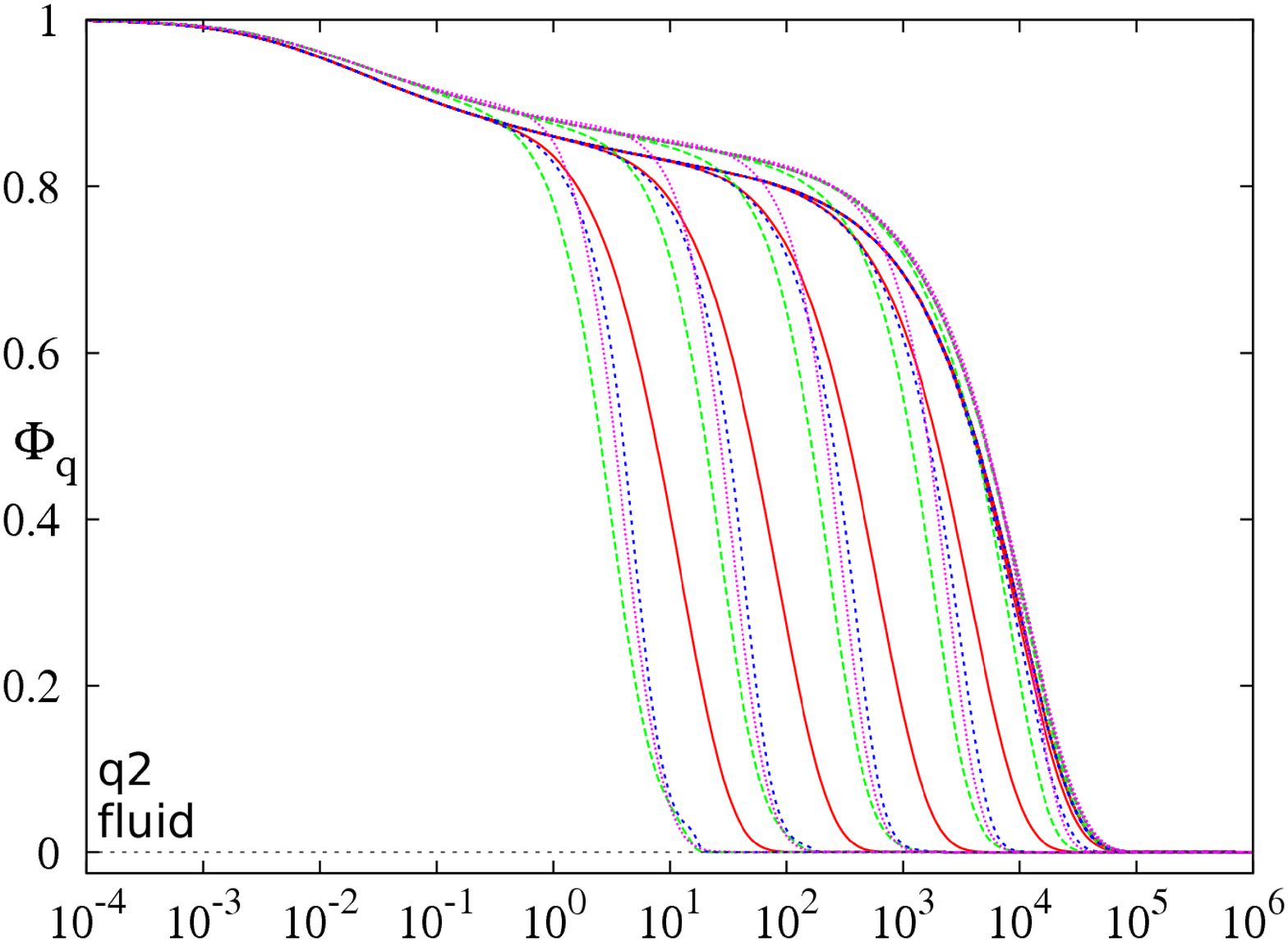}\hfill
 \includegraphics[clip = true, width=0.48\textwidth]{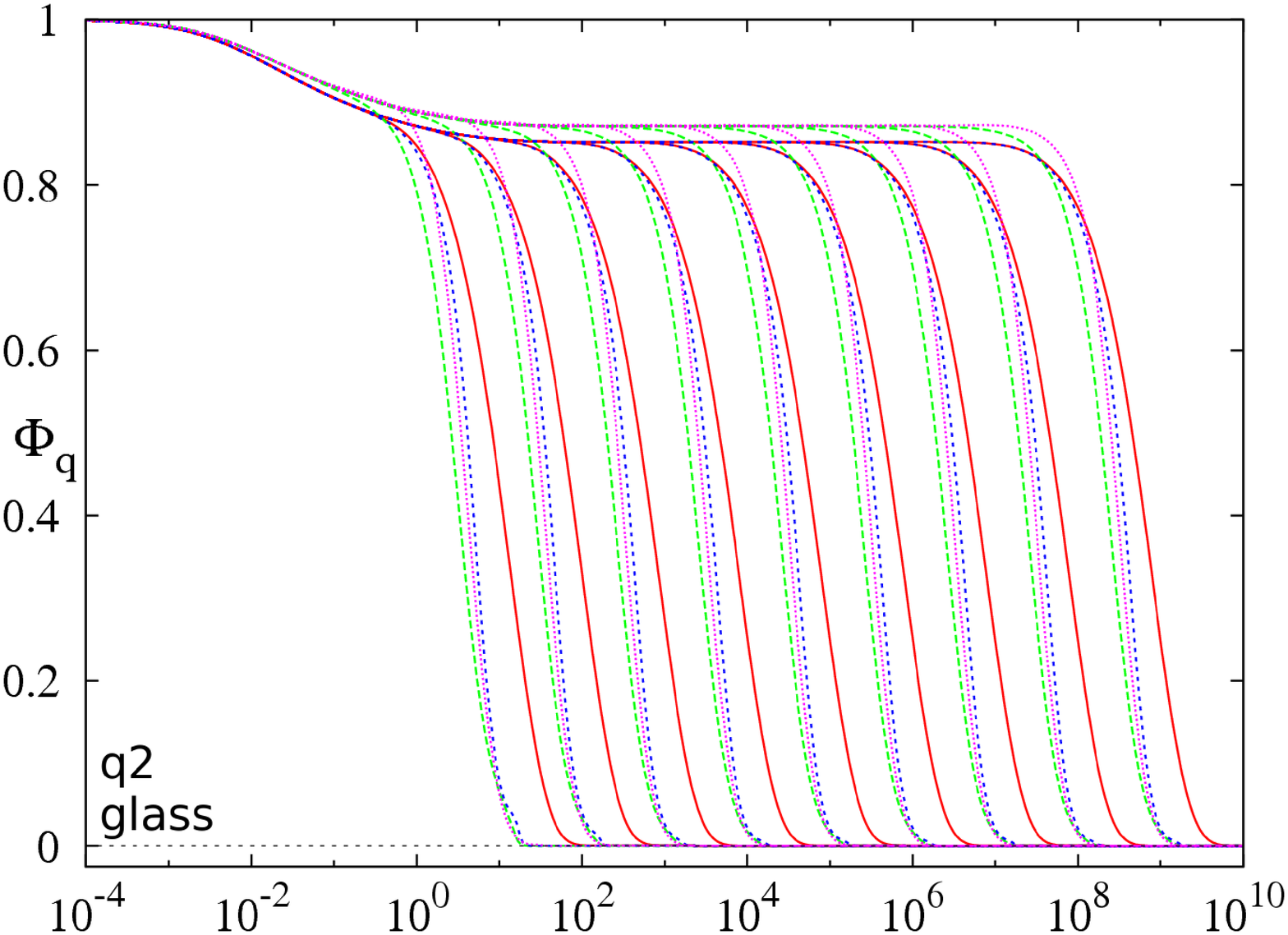}\\
 \includegraphics[clip = true, width=0.48\textwidth]{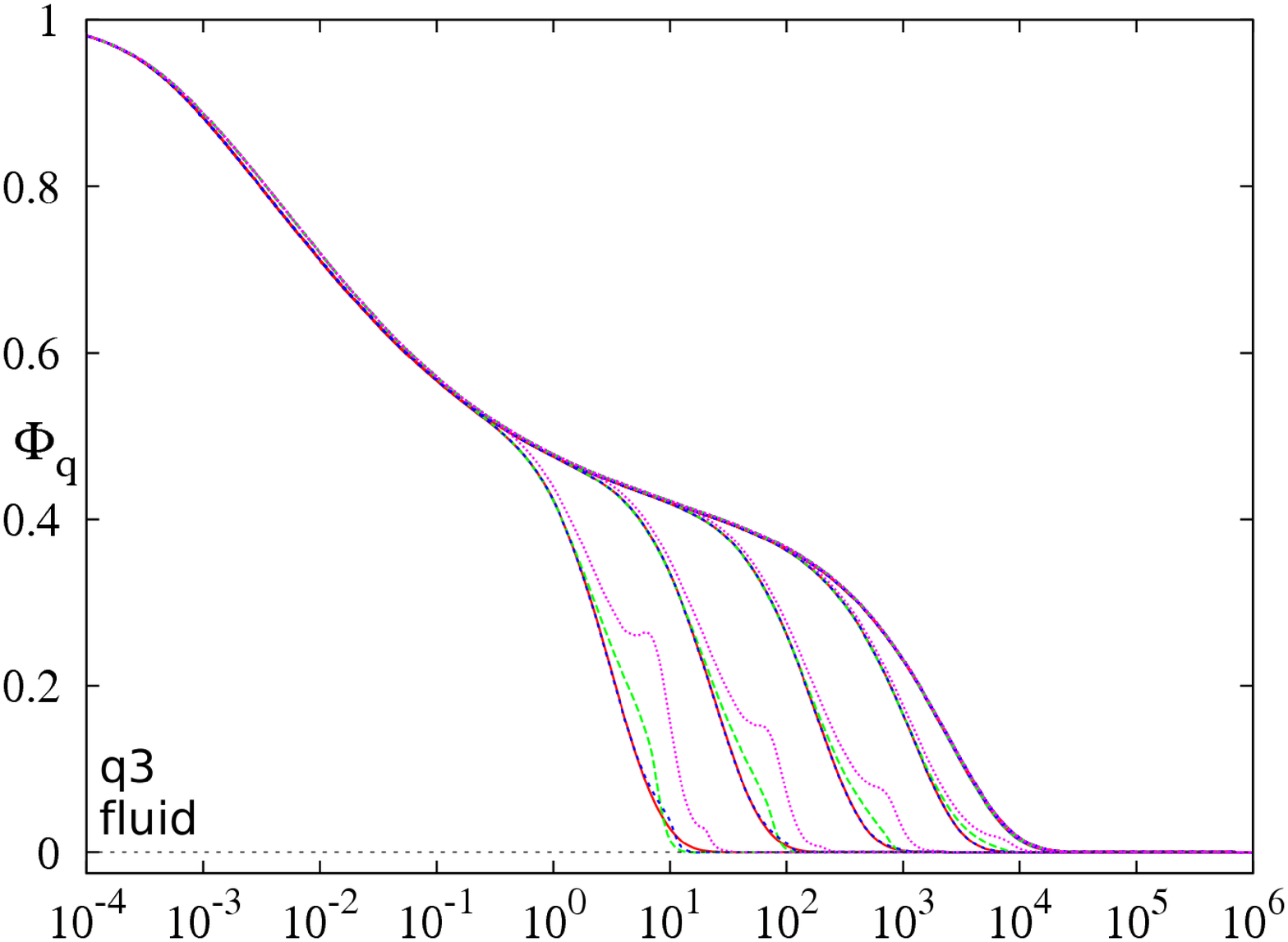}\hfill
 \includegraphics[clip = true, width=0.48\textwidth]{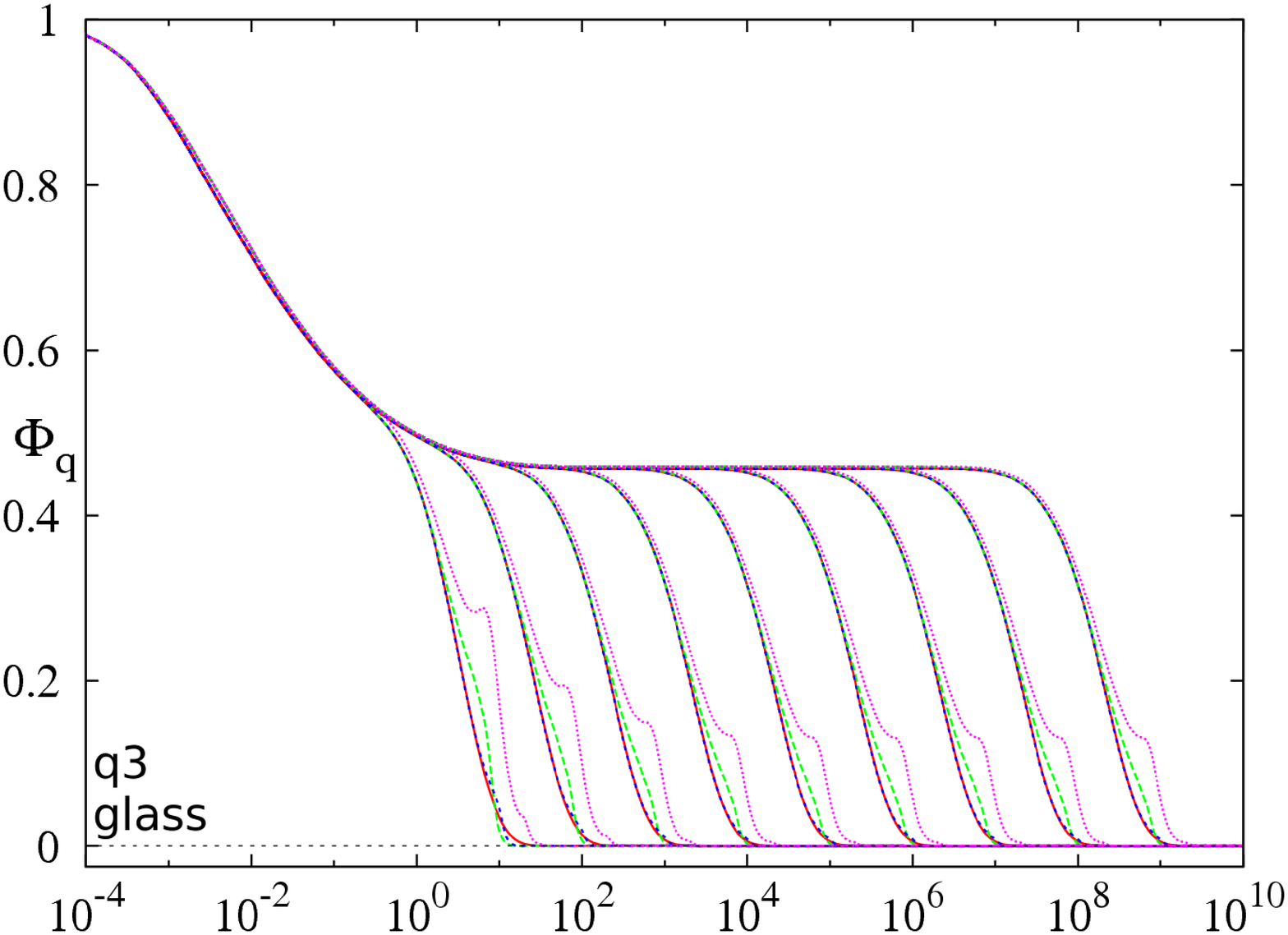}
 \includegraphics[clip = true, width=0.48\textwidth]{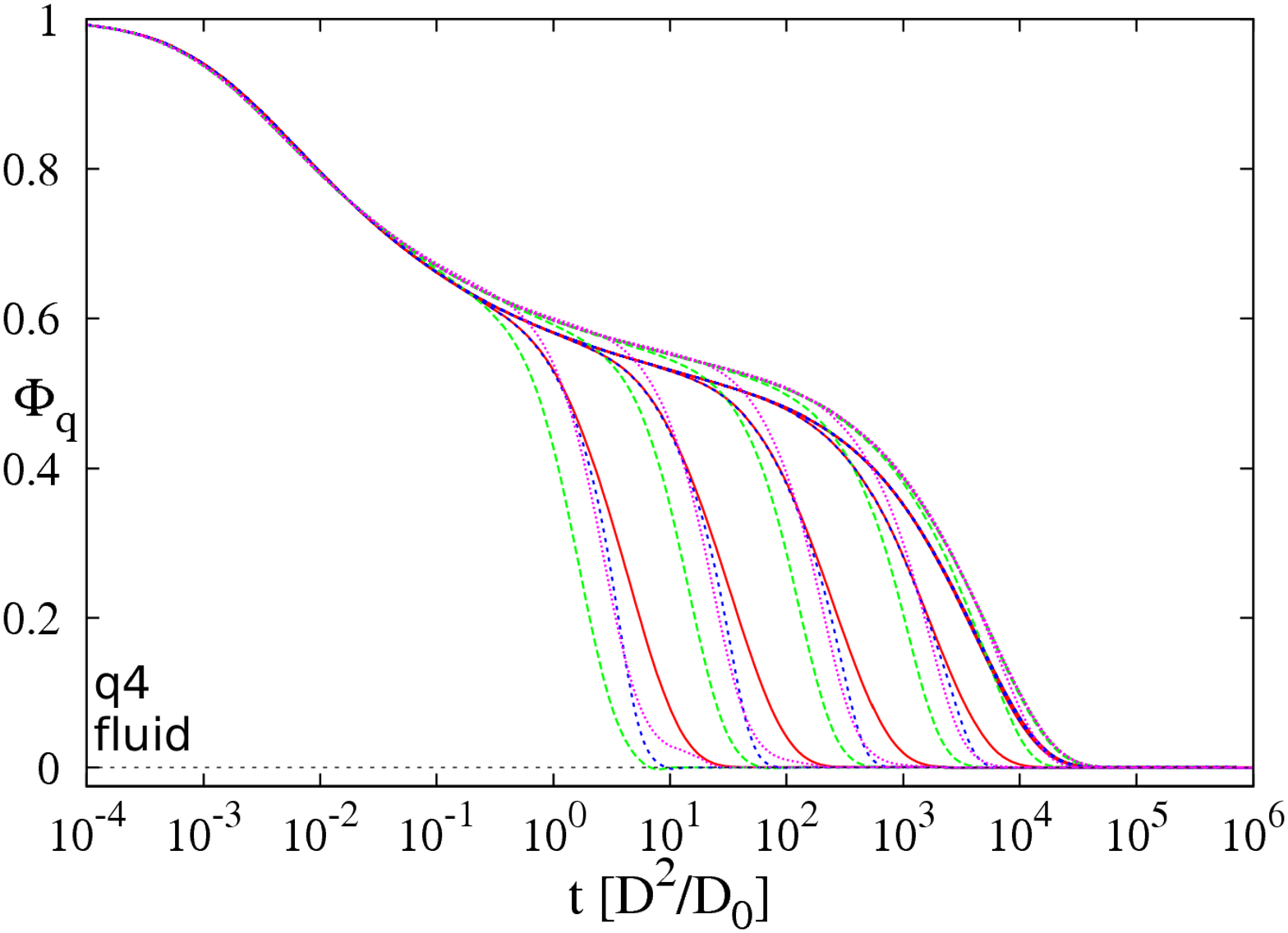}\hfill
 \includegraphics[clip = true, width=0.48\textwidth]{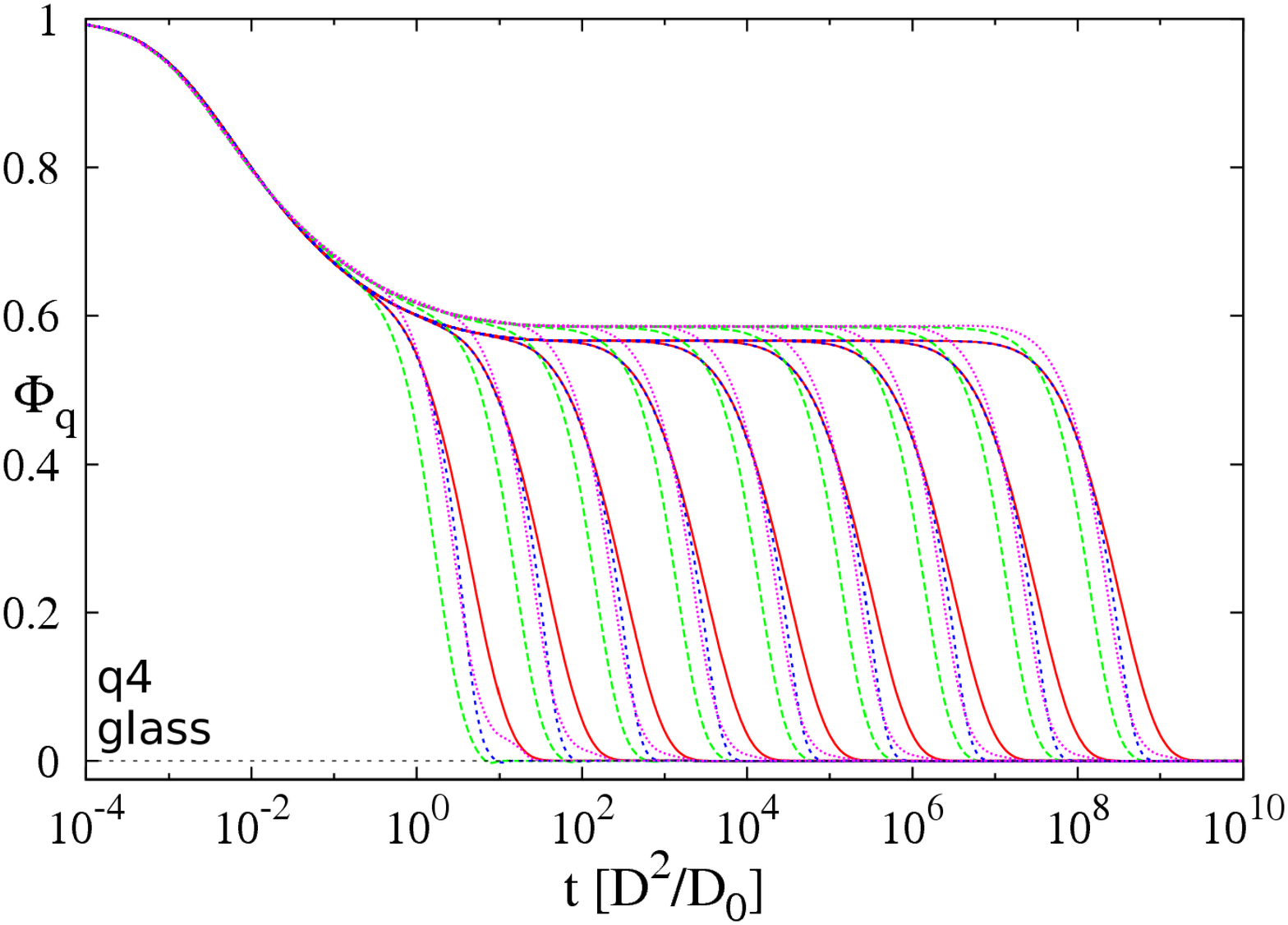}
\caption{Transient density
correlators in MCT in the liquid state (left column) for
$\epsilon=-10^{-3}$ and in the glassy state (right column) at
$\epsilon=10^{-3}$ at $q_1, q_2, q_3$ and $q_4$ (from top to bottom)
and Pe$_0=10^{-1}\dots 10^{-9}$ (from left to right). The colour
code gives the orientations according to Fig. \ref{fig1}. The
difference of the plateau heights in the second and fourth row
emerges as a discrepancy of the magnitudes and orientations of wavevectors due to the
selected discretization.}\label{fig2}
\end{figure}

The central quantities in MCT-ITT, which encode the competition
between shear driven motion and random fluctuations, are the
transient density correlators. Structural rearrangements manifest
themselves as a (second) slow relaxation process in the  $\Phi_{\bf
q}(t)$, whose relaxation time depends sensitively on the distance to
the glass transition, measured by the (relative) separation
parameter $\epsilon=(\phi-\phi_c)/\phi_c$, and on the magnitude of
the shear rate. Figure \ref{fig2} shows representative curves in a
fluid state (left column) and in a shear-melted glassy state (right
column) for the wavevectors and directions defined in Fig.
\ref{fig1}, and for nine different shear rates. The denoted bare
Peclet numbers measure the shear rate
compared to the Brownian diffusion time estimated with the single
particle diffusion coefficient $D_0$ at infinite dilution.
Characteristically, for the present strongly viscoelastic system,
shear affects the structural relaxation already at extremely small
bare Peclet numbers Pe$_0\ll1$. The short time motion, which
corresponds to the local diffusion of the particles within their
neighbour cages, however, is not much affected as long as Pe$_0<1$ holds.

In the fluid ($\epsilon<0$), a linear response regime, where shear
does not affect the decay of thermal fluctuations, is observed for
small (dressed) Peclet or Weissenberg numbers Pe$\;\le1$, where Pe$\;=
\dot\gamma \tau$ measures the shear rate relative to the intrinsic
relaxation time. The latter can be estimated from the
relaxation of $\Phi_{q_2}(t)$, viz. at the primary peak of $S_q$,
where the structure relaxes most slowly in the quiescent case. The
final (or $\alpha$-) relaxation time $\tau$ increases strongly when
approaching the glass transition; with
$\tau\propto(-\epsilon)^{2.38}$ predicted by MCT (Bayer {\it et al.} 2007). For
the fluid state in Fig. \ref{fig2}, only Pe$_0\;\le 10^{-6}$ is small
enough that Pe$\;<1$ holds and the final relaxation curves agree
for different shear rates. In  the glass ($\epsilon\ge0$),  shear is
the origin of the decay of the otherwise frozen-in density
fluctuations, and all nine shear rates lead to different final
decays. While for small and larger wavevectors the final decay is
rather isotropic, around the main peak in $S_q$ some anisotropy is
noticeable in the transient correlators. In the direction
perpendicular to the flow, $q_x=0$ (red), shear is not as efficient
in decorrelating the density as in the other directions, which fall
closer together.

\begin{figure}[htp]
\includegraphics[clip = true, width=0.48\textwidth]{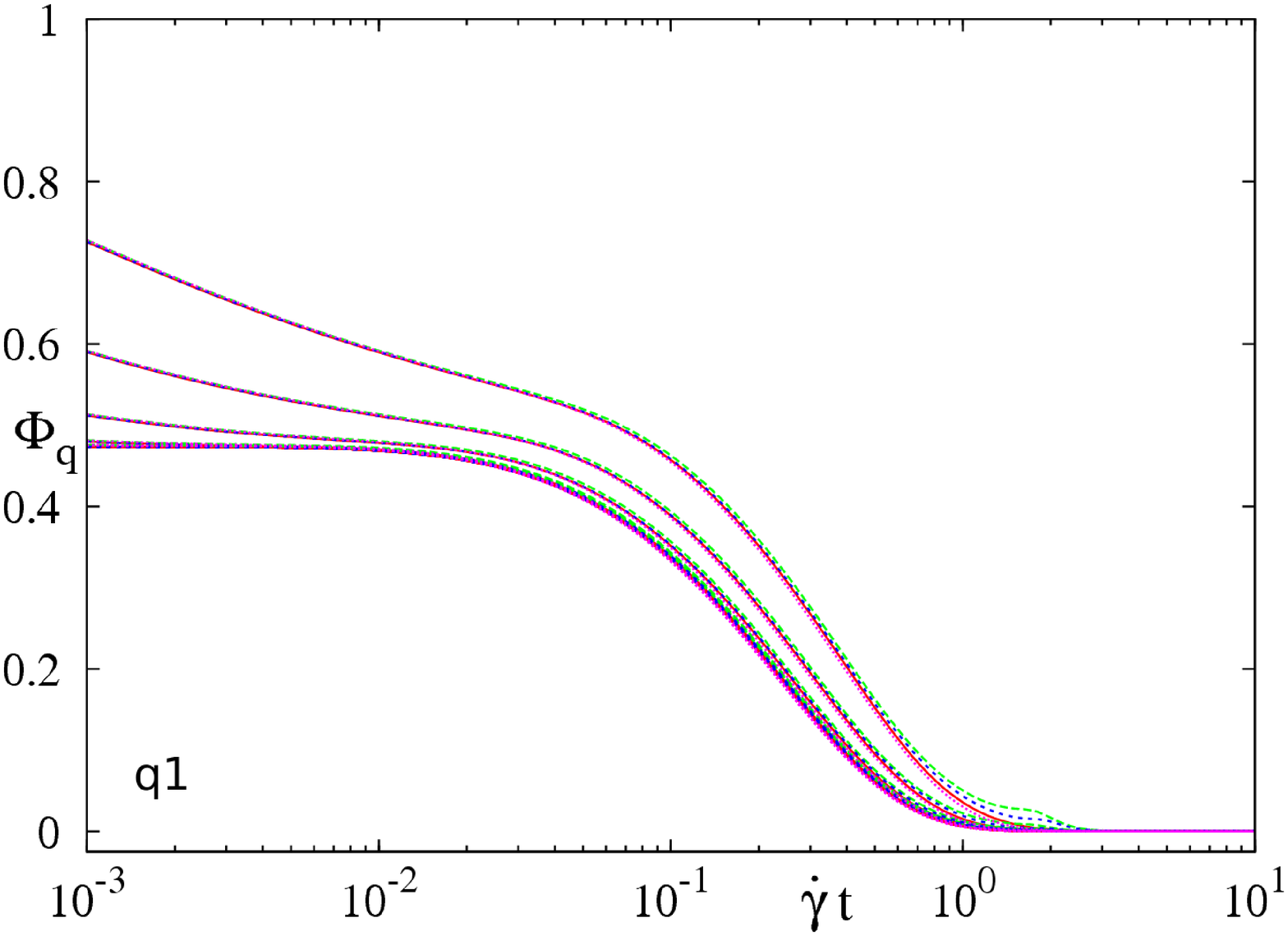}\hfill
\includegraphics[clip = true, width=0.48\textwidth]{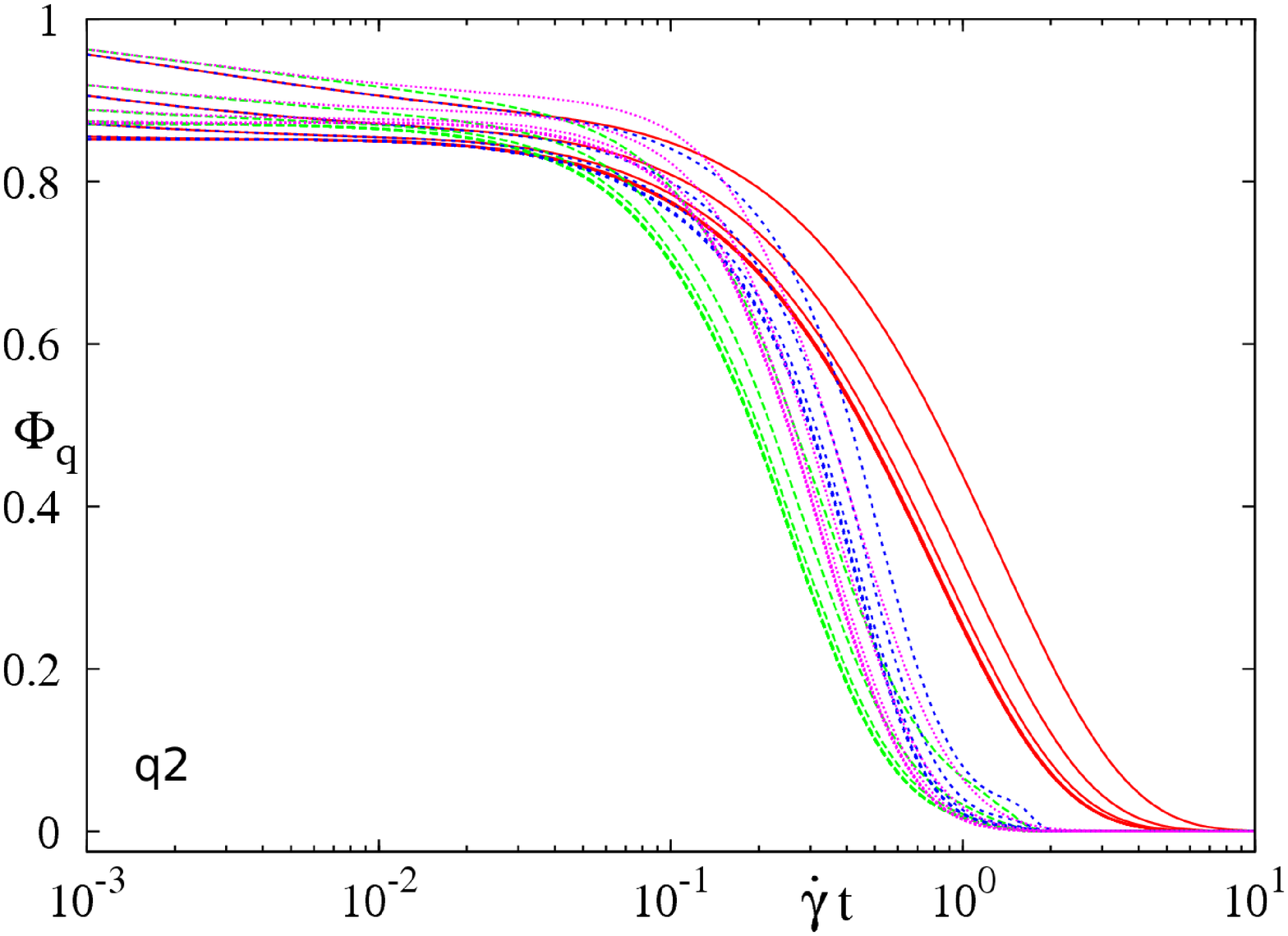}\\
\includegraphics[clip = true, width=0.48\textwidth]{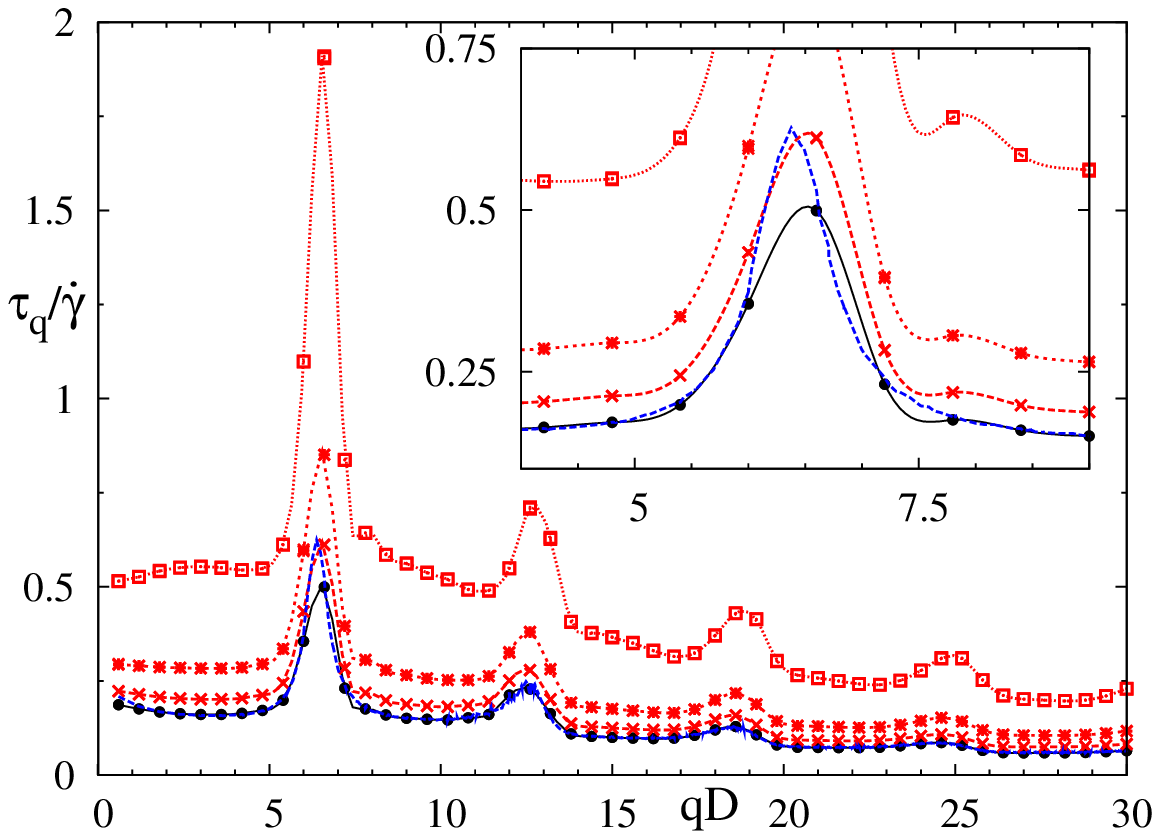}\hfill
\includegraphics[clip = true, width=0.48\textwidth]{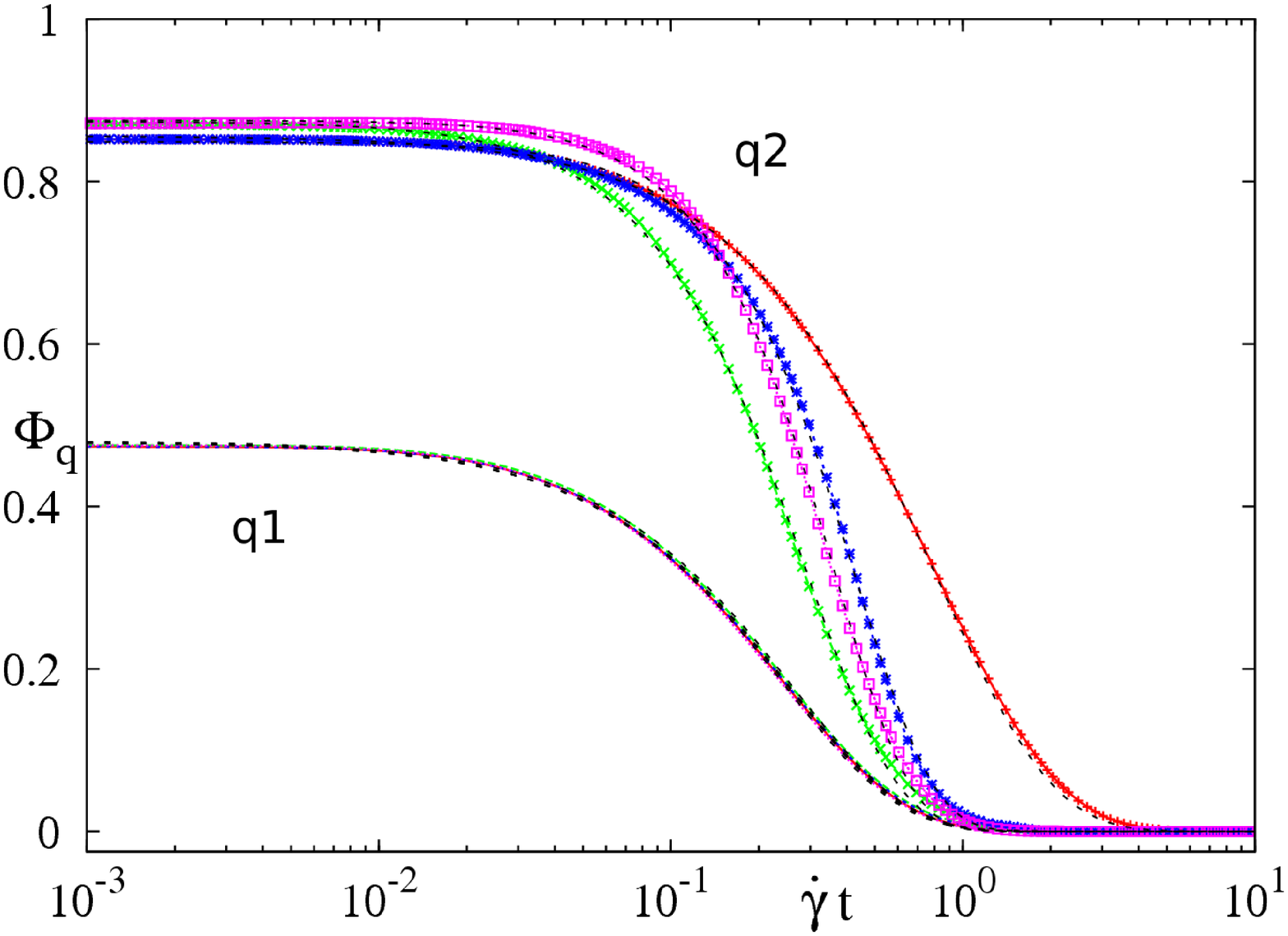}\\
\caption{Top row: transient density correlators
as functions of accumulated strain $ \dot{\gamma}t$ in the glassy state at
$\epsilon=10^{-3}$ for the wavevectors $q_1$ (top left), $q_2$ (top
right); collapse/ approach onto a master function appears for Peclet numbers ranging from Pe$_0=10^{-1}$ to $10^{-9}$. Bottom right, master functions of this glassy state at $q_1$ (isotropic, lines) and $q_2$ (anisotropic, differently colored symbols and lines).
The dashed black lines in this panel show compressed exponential fits
$\Phi_{\bf q}(t)\simeq A_{\bf q} \exp{(-(t/\tau_{\bf q})^{\beta_{\bf
q}})}$ with parameters according to Table \ref{table1}. Bottom left:
rescaled final relaxation times $\tau_{\bf q}/\dot{\gamma}$ for
$q_x=0$, Pe$_0=10^{-9}$ and $\epsilon=10^{-2}, 10^{-3}$ and
$10^{-4}$ (red from top to bottom) and $10^{-5}$ (black). The blue
line gives the estimate deduced from the stability analysis
$\tau_{\bf q}/\dot{\gamma}\simeq f_q^c/(h_q
\sqrt{c^{(\dot{\gamma})}/(\lambda-\frac{1}{2})})$ for
$c^{(\dot{\gamma})}=3.5$. The inset shows a close-up of the region
around the first peak.}\label{figyield}
\end{figure}

One of the central predictions of MCT-ITT concerns the existence of
a scaling law describing the yielding of glassy states,  which
manifests itself by an approach to a master function $\Phi_{\bf
q}(t) \to \tilde \Phi_{\bf q}(\tilde t)$ for decreasing shear rate,
$\dot\gamma\to0$ and $\epsilon\ge0$, where the rescaled time $\tilde t=\dot\gamma\, t$
agrees with the accumulated strain. This scaling law, which is quite
obvious in Fig. \ref{fig2}, is tested quantitatively for some wavevectors  in Fig. \ref{figyield}, and holds extremely well (for $q_1$), or with pre-asymptotic corrections (for $q_2$). The shapes of the yielding master functions can be very well fitted with
compressed exponentials,  $\Phi_{\bf q}(t)\simeq A_{\bf q}
\exp{(-(t/\tau_{\bf q})^{\beta_{\bf q}})}$, but the parameters,
including the exponent $\beta_{\bf q}$, depend on wavevector and
orientation; see Table \ref{table1} for representative values.

\begin{table}
\centering
\begin{tabular}{|| c | c | c | c | c | c | c ||}
\hline
${\bf q}$ & $|{\bf q}|/D$ & $q_x/D $ & $q_y/D$ & $A_{\bf q}$ & $\beta_{\bf q}$ & $\tau_{\bf q} \cdot \dot{\gamma}$ \\
\hline\hline
$q_1$ red & 3.0 & 0 & 3.0 & 0.479 & 1.121 & 0.255 \\
\hline
$q_1$ green & 3.0 & 2.4 & 1.8 &0.481 & 1.117 & 0.263 \\
\hline
$q_1$ blue & 3.0 & 3.0 & 0 &0.479 & 1.112 & 0.258 \\
\hline
$q_1$ magenta & 3.0  & 1.8 & -2.4 & 0.480 & 1.117 & 0.249 \\
\hline
$q_2$ red & 6.6 & 0 & 6.6 & 0.856 & 1.084 & 0.811\\
\hline
$q_2$ green & 6.5 & 5.4 & 3.6 & 0.878 & 1.384 & 0.291\\
\hline
$q_2$ blue & 6.6 & 6.6 & 0 & 0.875 & 1.670 & 0.363\\
\hline
$q_2$ magenta & 6.5 & 3.6 & -5.4 & 0.849 & 1.617 & 0.431 \\
\hline\hline
\end{tabular}
\caption{Parameters of the compressed exponentials $A_{\bf q}
\exp{(-(t/\tau_{\bf q})^{\beta_{\bf q}})}$ fitted to the yield
master functions  shown in Fig. \ref{figyield}.} \label{table1}
\end{table}

The quite good fit to compressed exponentials is at present only a
numerical observation. Yet, close to the glass transition asymptotic
expansions are possible which analytically predict the initial part
of the yield master functions (Fuchs \and Cates 2002,2003,2009):
$$\tilde \Phi_{\bf q}(\tilde t\to0,\epsilon=0) \to f^c_q - h_q \, \sqrt{c^{(\dot{\gamma})}/(\lambda-\frac{1}{2})}) \; \tilde t +
\ldots \; .$$ As the critical glass form factor $f^c_q$ and the
critical amplitude $h_q$ are isotropic, this result suggests a
rather isotropic yielding process right at the glass transition
density. Approximating $\tilde \Phi_{\bf q}(\tilde t)$ by an
exponential, it also provides an estimate for the final relaxation
time under shear. Except for $c^{(\dot{\gamma})}$ all quantities in
the above formula have been determined in quiescent MCT (Bayer {\it et al.} 2007),
and our values (e.g. $\lambda=0.698$) differ only because
of the coarser discretization in $\bf q$ space that is necessary
under shear. Close to the glass transition, we estimate
$c^{(\dot{\gamma})}\approx 3.5$. As the bottom left panel in Fig.
\ref{figyield} and Fig. \ref{fig2} both show, this estimate for
the final relaxation time is quite good close to the transition, but
deeper in the glass the correlators become somewhat more
anisotropic. Specifically for $q_x=0$ (red) with wavevector magnitudes
around $|{\bf q}|\simeq q_2$, the correlator slows down relative to other orientations.

\begin{figure}[htp]
\includegraphics[clip = true, width=0.48\textwidth]{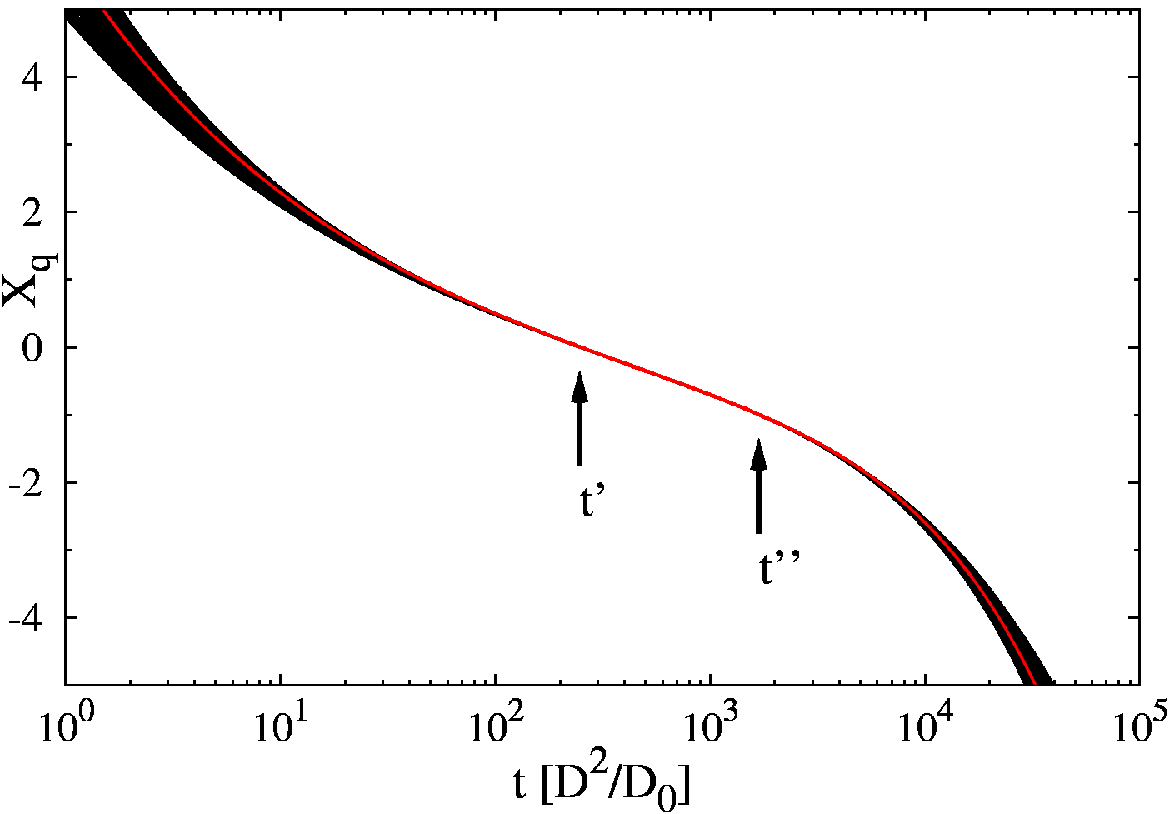}\hfill
\includegraphics[clip = true, width=0.48\textwidth]{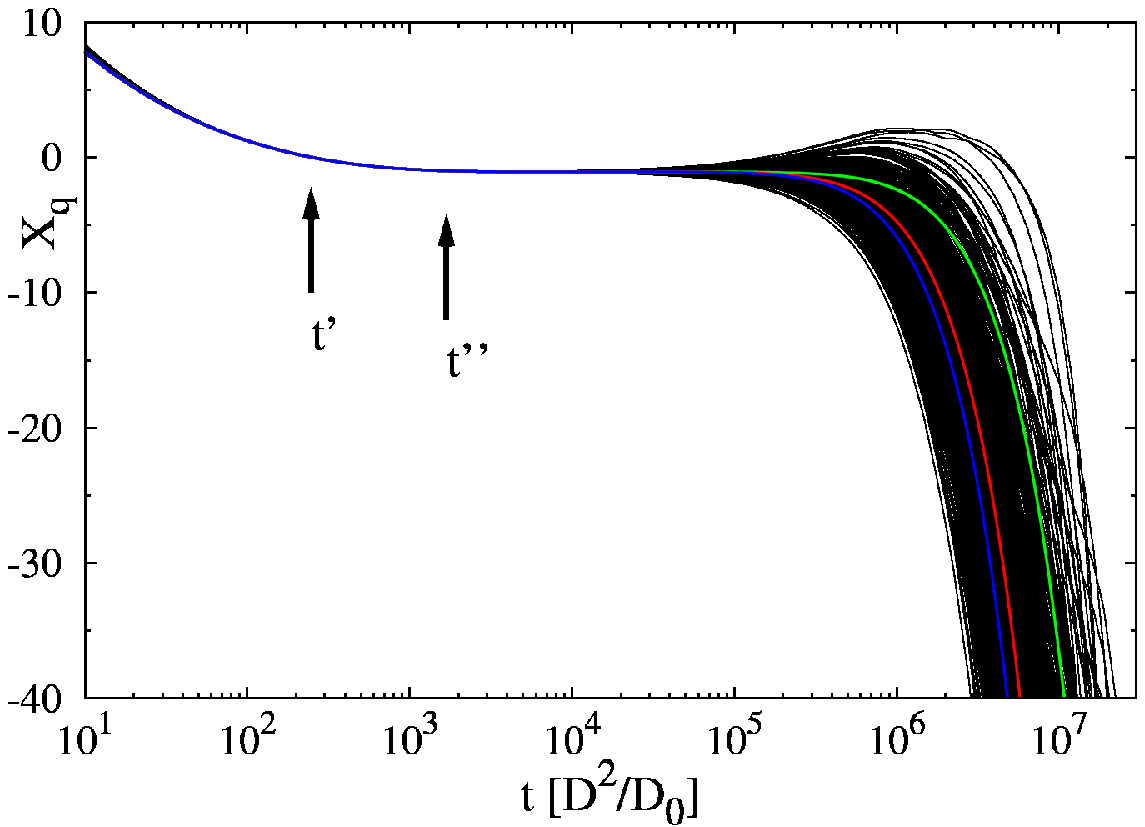}\hfill
\caption{Factorization property: $X_{\bf q}(t)=\frac{\Phi_{\bf
q}(t)-\Phi_{\bf q}(t')}{\Phi_{\bf q}(t')-\Phi_{\bf q}(t'')}$. The
left picture shows the liquid state at $\epsilon=-10^{-4},
\dot{\gamma}=10^{-8}$, while the right picture depicts the glassy
state at $\epsilon=10^{-4}, \dot{\gamma}=10^{-8}$. Black curves give
the solutions of the MCT equations for $0\le qD\le 24$. At the times
$t'=247$ and $t''=1683$  the correlators are
approximately $0.01$ above, respectively below their critical
plateaus. Coloured curves are rescaled solutions of the modified
$\beta$-scaling equation
$X(t)=\frac{{\cal G}(t/t_\sigma)-{\cal G}(t'/t_\sigma)}{{\cal G}(t'/t_\sigma)-{\cal G}(t''/t_\sigma)}$
with $t_\sigma=965$ and correspond to values $c^{(\dot{\gamma})}=1$ (green), $3.5$ (red) and $5$ (blue). The coefficient $c^{(\dot{\gamma})}=3.5$
was also found in Fig. \ref{figyield} and captures the shear driven decay of
the glass structure.}\label{figbeta}
\end{figure}

More detailed analytical predictions, are possible around $f^c_q$,
where  spatial and temporal dependences in the transient
fluctuations decouple (Fuchs \and Cates 2003):
$$\Phi_{\bf q}(t) = f^c_q + h_q \;
\sqrt{C |\epsilon|}\; {\cal
G}_\pm(t/t_\sigma, |\dot\gamma|\,t_\sigma \sqrt{c^{(\dot{\gamma})} / C |\epsilon|})
+ \ldots\; .$$ ${\cal G}_\pm$ is a universal function, which only
depends on $\lambda$, and $t_\sigma\propto |\epsilon|^{-1.56}$ is
another MCT time scale that diverges at the glass transition
(Bayer {\it et al.} 2007). This factorization property known from quiescent MCT
generalizes to steady shear, and is an
essential feature of the localization transition that underlies
glass formation in MCT. In Fig. \ref{figbeta}, the quantity $X_{\bf
q}(t)=\frac{\Phi_{\bf q}(t)-\Phi_{\bf q}(t')}{\Phi_{\bf
q}(t')-\Phi_{\bf q}(t'')}$ is plotted which should become wavevector
and orientation independent if factorisation holds. This holds in
the liquid, and in the glass, where however shear leads to strong
(anisotropic) pre-asymptotic corrections already  for $\dot\gamma \,
t \ge 0.01$.

\section{Results and comparisons}

Based on the approximated generalized  Green-Kubo relations of Eqs.
(\ref{sigma_xy}) and (\ref{distsq}), and the properties of the
transient correlators discussed in the previous section, MCT-ITT
makes a number of predictions for stationary stresses and structural
correlations (Fuchs \and Cates 2002, 2009). In the following, we will test
these qualitatively, but also quantitatively, by comparing them to
the two-dimensional simulation data.

\subsection{Stationary stresses}

\begin{figure}[htp]
\includegraphics[clip = true, width=\textwidth]{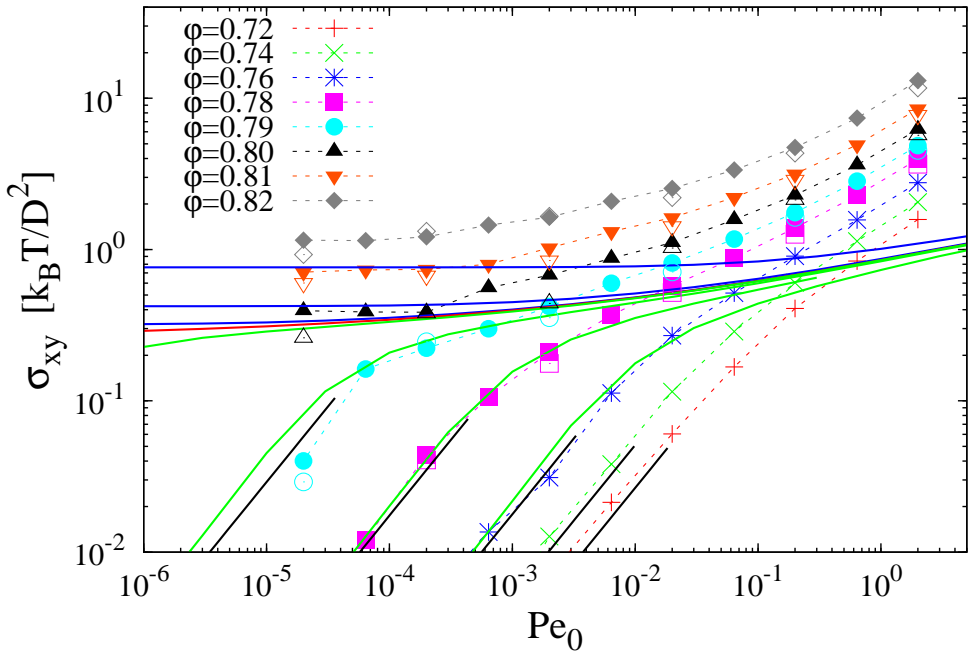}\\
\caption{Shear stress $\sigma_{xy}(\dot\gamma)$ versus shear rate given as Peclet number Pe$_0=\dot\gamma D^2/D_0$ in
MCT-ITT and simulation: The data points give simulation results for the densities denoted in the legend. Empty symbols show the results obtained via (\ref{contactvalue}) and filled symbols the results from (\ref{coll_sigma}). The Pe$_0 \to 0$ (black solid lines) were calculated according to (\ref{shearviscosity}).
The solid lines show  calculations in MCT-ITT for $\epsilon=10^{-2}, 10^{-3}, 10^{-4}, 10^{-5}, -10^{-4}, -10^{-3}$ and $-10^{-2}$ (from top to bottom; blue $\epsilon>0$, red $\epsilon\approx 0$, green $\epsilon<0$).  MCT-ITT results are shifted downwards by a factor 0.1 to match the simulation results. }\label{figstress}
\end{figure}

The quantity of most interest in nonlinear rheology is the shear
stress $\sigma_{xy}(\dot\gamma)$. 
Flow curves giving the shear stress as function of the
shear rate can be obtained in the simulations and are shown in Fig.
\ref{figstress}. The viscosity
decreases below its Newtonian value
$\eta_0=\sigma_{xy}(\dot\gamma\to0)/\dot\gamma$  upon increasing the
shear rate already for small Pe$_0$ values. Shear
thinning in which the stress increases less than linearly with
$\dot\gamma$, sets in at Pe of the order of unity, and this crossover
shifts to lower and lower Pe$_0$ for increasing density. At the
density around $\phi_c\approx 0.79$, the crossover leaves the
accessible shear rate window. We use this as estimate of an ideal
glass transition density, where the final relaxation time and the
quiescent Newtonian viscosity ($\eta_0\propto\tau$) diverge.
Moreover, at this density the flow curves change from a
characteristic S shape in the fluid, to exhibiting a rather
$\dot\gamma$-independent plateau at small shear rates. This change
is the hallmark of the transition in MCT-ITT between a shear
thinning fluid and a yielding glass (Siebenb\"urger {\it et al.} 2009).

\begin{figure}[htp]
\includegraphics[clip = true, width=\textwidth]{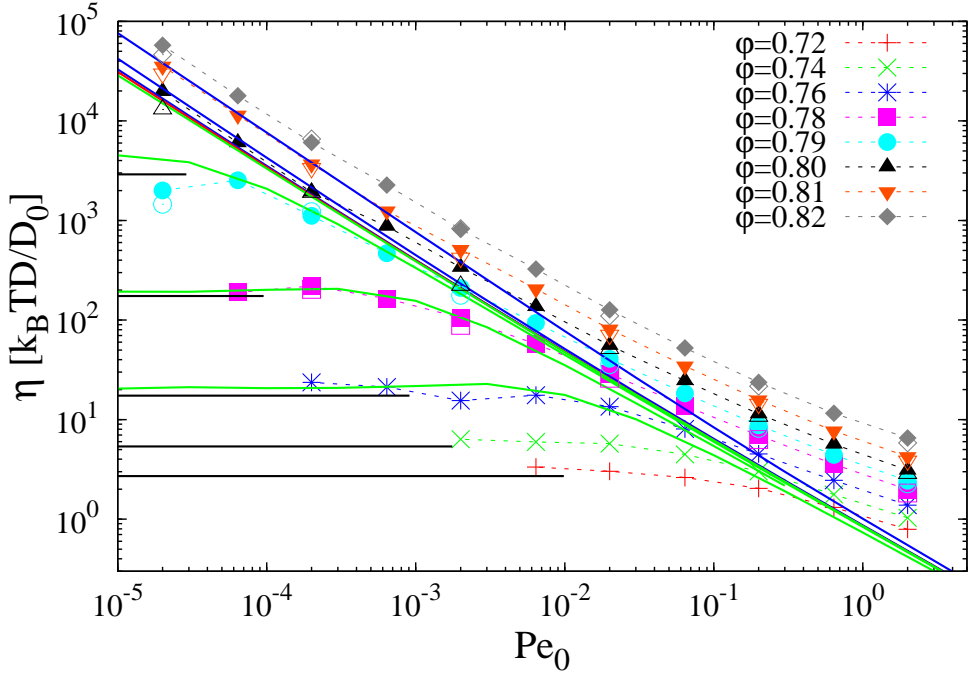}
\caption{Viscosity $\eta (\dot \gamma)$ versus shear rate given as Peclet number Pe$_0=\dot\gamma D^2/D_0$ in MCT-ITT and simulation. The color codes are
the same as in Figure \ref{figstress}. Again the empty symbols show the results of (\ref{contactvalue}) and filled symbols the results from (\ref{coll_sigma}) while the Pe$_0 \to 0$ (black solid lines) were calculated according to (\ref{shearviscosity}).  MCT-ITT results are shifted downwards by a factor 0.1 to match the simulation results.}\label{figeta}
\end{figure}
The numerical MCT-ITT solutions show the same transition scenario,
but because of the approximations involved, exhibit a different
critical density than the simulations. Even if theory and simulation
considered the same (binary) system, a difference in critical
packing fraction would be expected as is well known also in three dimensions (Voigtmann {\it et al.} 2004). Thus the following MCT-ITT
calculations match the relative separation $\epsilon$ from $\phi_c$ in
order to compare with the simulations. As MCT-ITT is aimed at describing the long time structural motion, errors need to be anticipated in its description of short time properties.  This is obvious in real dispersions, where hydrodynamic interactions (neglected in MCT-ITT) affect the short time diffusion coefficient $D_s$, but  may arise in the following comparisons, also. A rescaling of the effective Peclet number Pe$_0^{\rm eff}=\dot\gamma D^2/D_s$ would correct for this change in $D_s$. In order not to introduce additional fit parameters we refrain from doing so, but anticipate that future comparisons may require $D_s\ne D_0$. 

The approach to a yield scaling law, where the final decay of the
transient correlators depends on the accumulated strain only (see
Figs. \ref{fig2} and \ref{figyield}), predicts the existence of a
(dynamical) yield stress
$\sigma_{xy}^+(\epsilon)=\sigma_{xy}(\dot\gamma\to0,\epsilon)$, which
characterizes the shear melted glass. In the bidisperse hard disc mixture, at the glass transition it takes the (critical) value $\sigma_{xy}^{+,c}=\sigma_{xy}^+(\epsilon=0)\approx 0.3 k_BT/D^2$; below the glass transition, the yield stress jumps to zero, $\sigma_{\alpha\beta}^+(\epsilon<0)=0$. Its quantitative prediction is
quite a challenge for theory, because Eq.~\ref{sigma_xy} shows that
it requires an accurate calculation of the shear driven relaxation
process. MCT-ITT overestimates the critical yield stress
$\sigma_{xy}^{+,c}$ by roughly a factor 10 because,
presumably, the decay of the transient correlators is too slow. Yet,
of course the difference between the monodisperse system in the
MCT-ITT calculation and the bidisperse simulated system contributes
in unknown way to the error. It appears reasonable to
assume that mixing two species reduces the stresses under flow,
which would explain part of the deviation.\\
At larger shear rates, the flow curves from simulation appear to approach a second Newtonian
plateau which, presumably, strongly depends on
the hard-core character of excluded volume interactions and is outside the reach of the
present MCT-ITT. The latter, by using its sole input $S_q$ rather than pair potential, is not directly aware of hard-core constraints.
We checked, however, that the states remain
homogeneous and random up to the Pe$_0$ values shown.

\begin{figure}[htp]
\includegraphics[clip = true, width=0.48\textwidth]{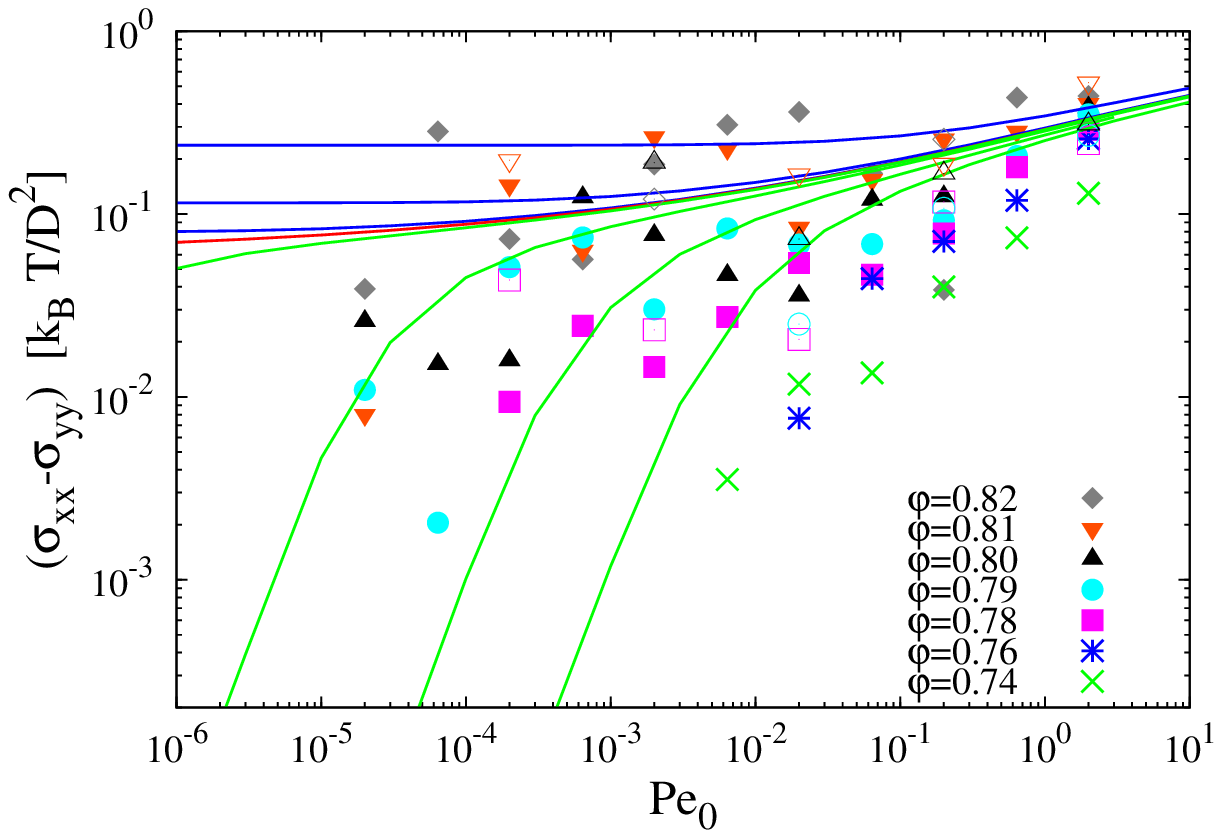}\hfill
\includegraphics[clip = true, width=0.48\textwidth]{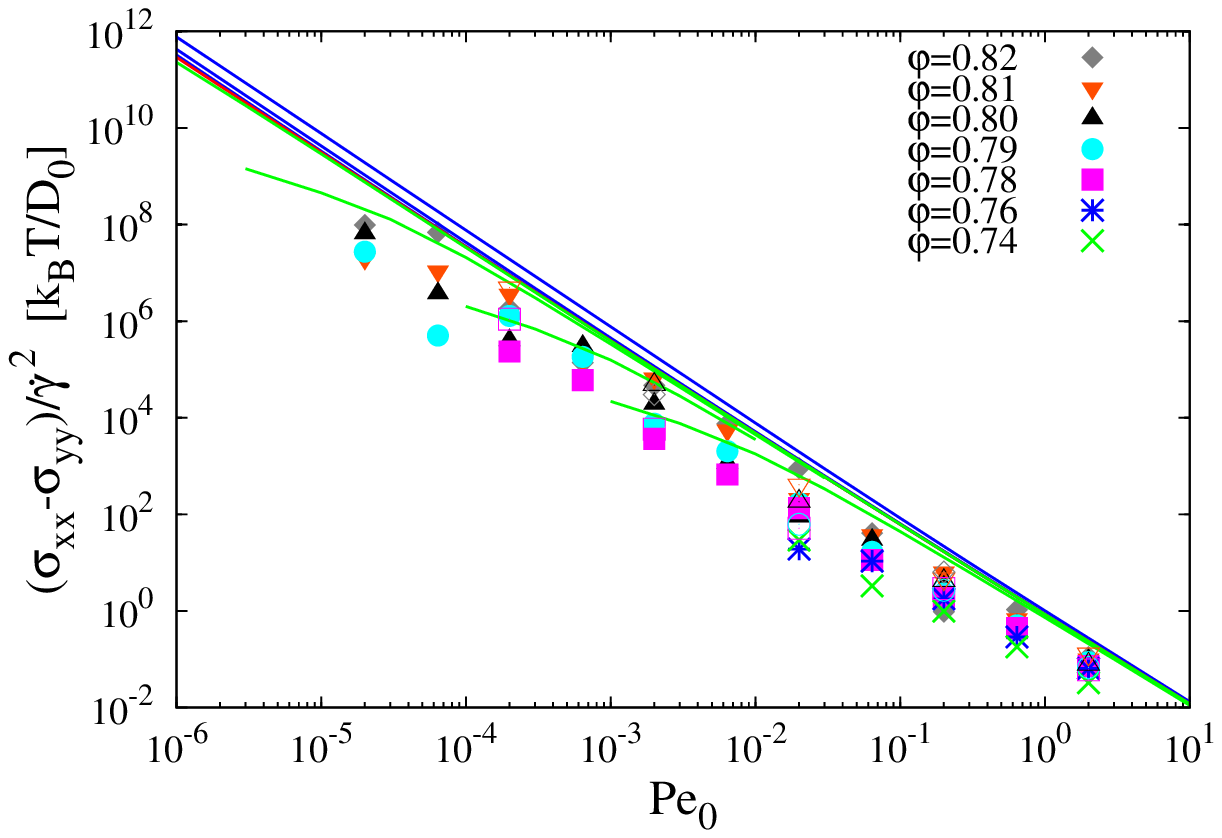}
\caption{Left panel: Normal stress differences for the system at the densities given in the legend. The lines show MCT-ITT results at separation parameters $\epsilon=10^{-2}, 10^{-3}, 10^{-4}, 10^{-5}, -10^{-4}, -10^{-3}$ and $-10^{-2}$ (from top to bottom) scaled by a factor 0.1 to match the simulation results. Right panel: Normal stress differences for the system divided by Pe$_0^2$. Color codes and scaling factors are the same as in the left panel. In both panels empty symbols show the results obtained via (\ref{contactvalue}) while filled symbols were calculated with (\ref{coll_sigma}). }\label{fignorm}
\end{figure}

Reassuringly, the same rescaling factor of 0.1 as for the shear
stress brings theoretical and simulational normal stress
differences, $\sigma_{xx}-\sigma_{yy}$, into register also; see Fig.
\ref{fignorm}. The normal stress differences are positive
(indicating that the dispersion would swell after flowing through a nozzle), and
show similar behavior to the stress, increasing like
$\dot\gamma^2$ in the fluid for small shear rates, and leveling out
onto a plateau in the yielding glass.

\subsection{Distorted micro-structure}

The macroscopic stresses in the flowing dispersion are experimentally most important, but provide only an averaged description of the local effects of shear. Spatially resolved information can be learned from the distorted structure factor $\delta S_{\bf q}(\dot\gamma)=S_{\bf q}(\dot\gamma)-S_q$, which in MCT-ITT is connected to the stress via 
$$
\sigma_{ij}= \frac{n\, k_BT}{2}\; \int \!\!\frac{d^2 q}{(2\pi)^2}\; 
\frac{q_i q_j}{q}\, \frac{\partial c_q}{\partial q} \;  \delta S_{\bf q}(\dot\gamma)\; ,
$$
as follows from Eqs.~(\ref{sigma_xy}) and (\ref{distsq}).
 
\begin{figure}[htp]
\includegraphics[clip = true, width=0.46\textwidth]{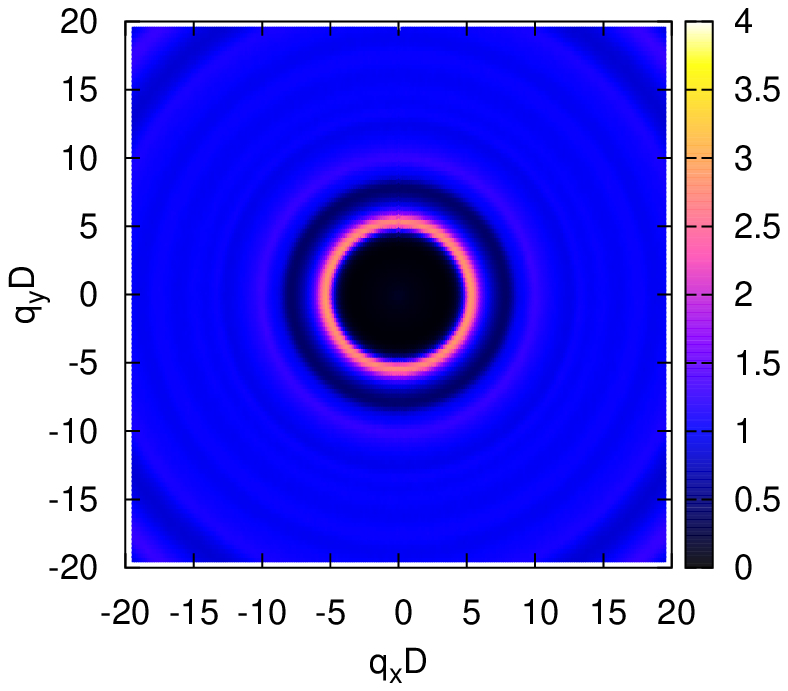}\hfill
\includegraphics[clip = true, width=0.46\textwidth]{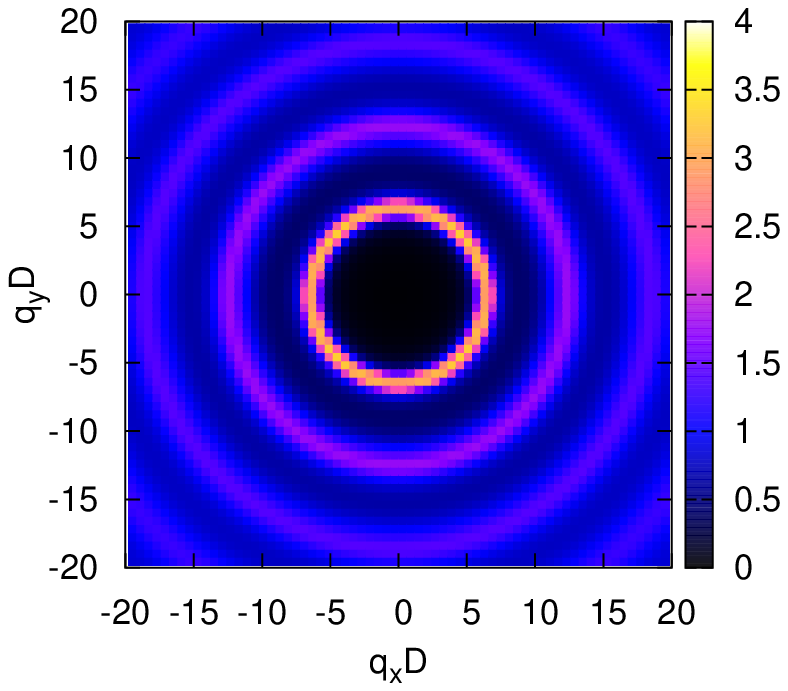}\\
\includegraphics[clip = true, width=0.46\textwidth]{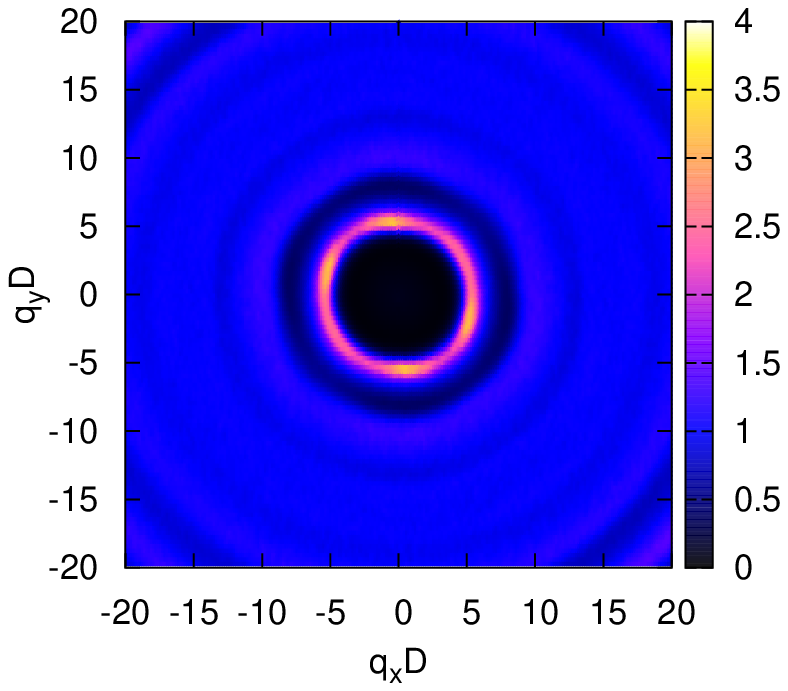}\hfill
\includegraphics[clip = true, width=0.46\textwidth]{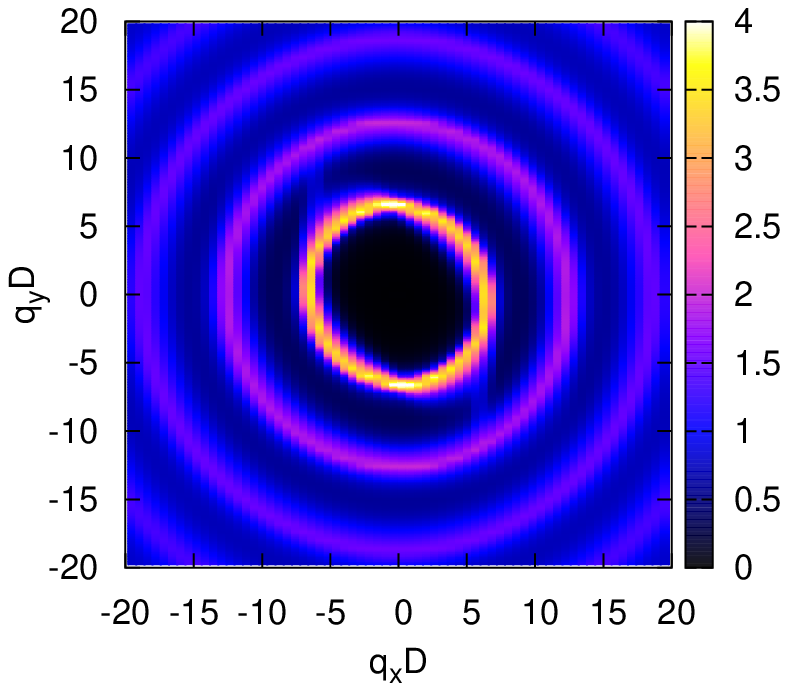}\\
\includegraphics[clip = true, width=0.46\textwidth]{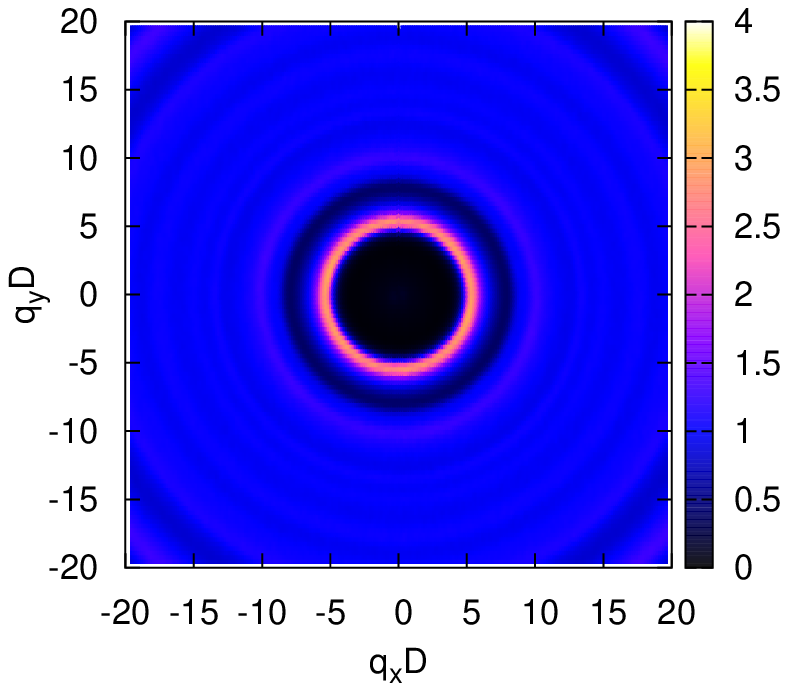}\hfill
\includegraphics[clip = true, width=0.46\textwidth]{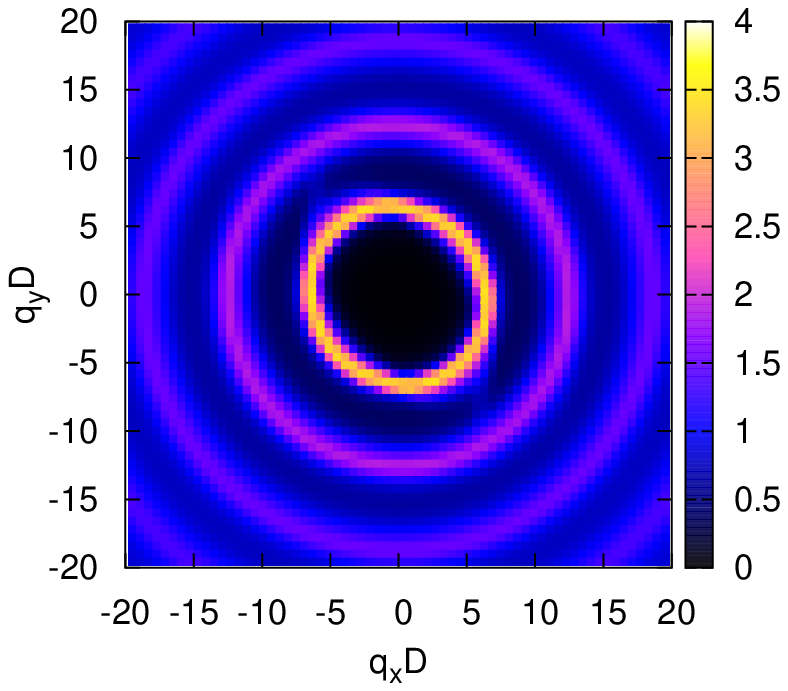}
\caption{Distorted structure factor $S({\bf q},\dot{\gamma})$ from
simulation (left column) and from theory (right column): for states
in linear response regime in the liquid (top row: simulation $\varphi =0.78$ and Pe$_0 = 2 \cdot 10^{-4}$ 
MCT $\epsilon=-10^{-2}$ and
Pe$_0=10^{-4}$); in the glass at high shear rate (middle row: simulation
$\varphi =0.79$ and Pe$_0= 2$,  MCT $\epsilon=10^{-3}$ and
Pe$_0=10^{-2}$); and  in the glass at small shear rate (bottom row:
simulation $\varphi =0.79$ and Pe$_0= 2 \cdot 10^{-4}$,  MCT
$\epsilon=10^{-3}$ and Pe$_0=10^{-4}$).}\label{figsqcolor}
\end{figure}

Figure \ref{figsqcolor} shows colour-coded structure factors $S_{\bf q}(\dot\gamma)$ as function of the two-dimensional wavevector $\bf q$, with $q_x$ in the direction of flow and $q_y$ along the gradient direction. In the left column, panels with simulation data are compared to panels in the right column  obtained in MCT-ITT. Scattering intensities are presented $(i)$ (top row) in the linear response regime in the fluid, effectively measuring the equilibrium structure factors already shown in Fig. \ref{fig1}, $(ii)$ (middle row) at high shear in the glass, where all densities are in the shear thinning region, and  $(iii)$ (bottom row) in the glass at low shear rate, where the yielding glassy state is tested. While the fluid $S_{\bf q}(\dot\gamma)$ is isotropic for small Pe$_0$ (case $(i)$), as required by linear response theory, increasing Pe$_0$ to values around unity, leads to an ellipsoidal scattering ring, which is elongated along the so-called 'compressional axis' $q_x=-q_y$, and more narrow along the 'extensional axis' $q_x=q_y$; this indicates that shear pushes particles together along the compressional and pulls particles apart along the extensional diagonals (Vermant  \and Solomon 2005). Theory and simulation data in this representation qualitatively agree except for that MCT-ITT somewhat overestimates the anisotropic distortion of the glass structure at low Pe$_0$.

\begin{figure}[htp]
\includegraphics[ width=0.46\textwidth]{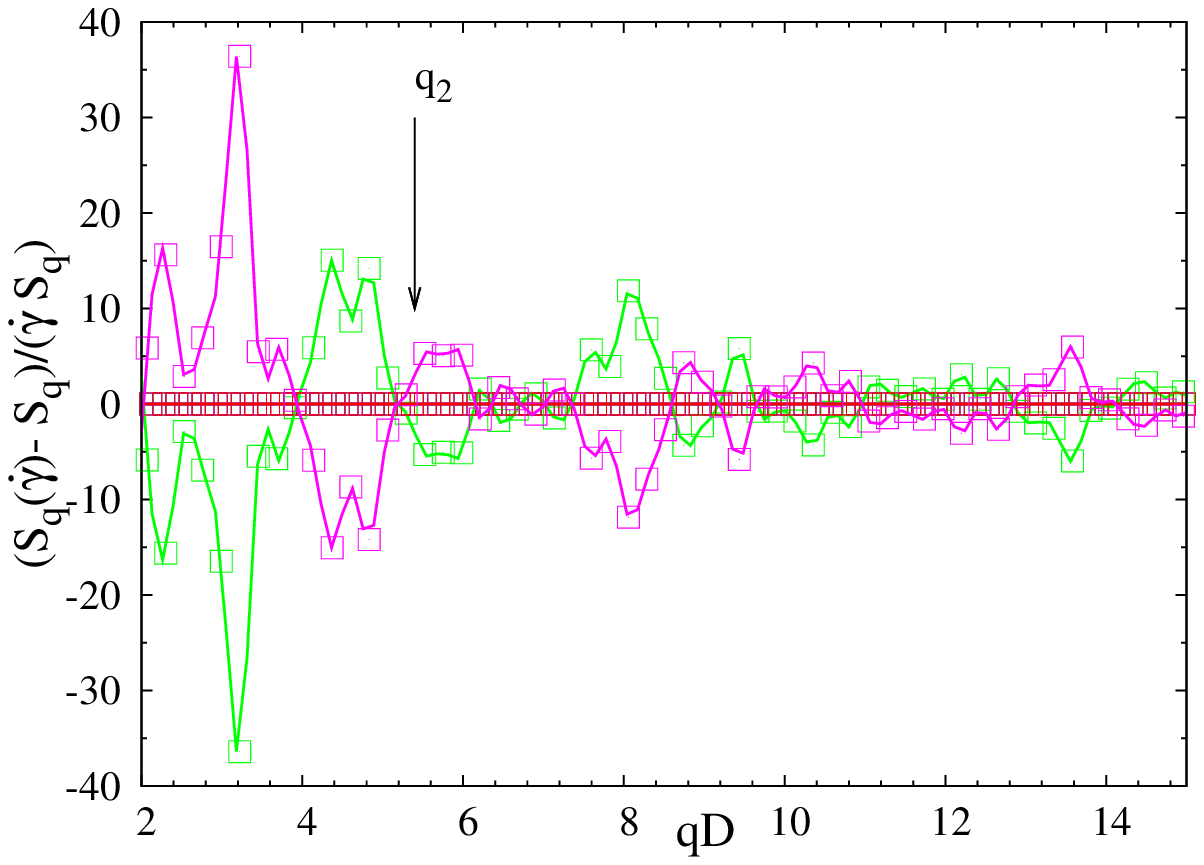}\hfill
\includegraphics[ width=0.46\textwidth]{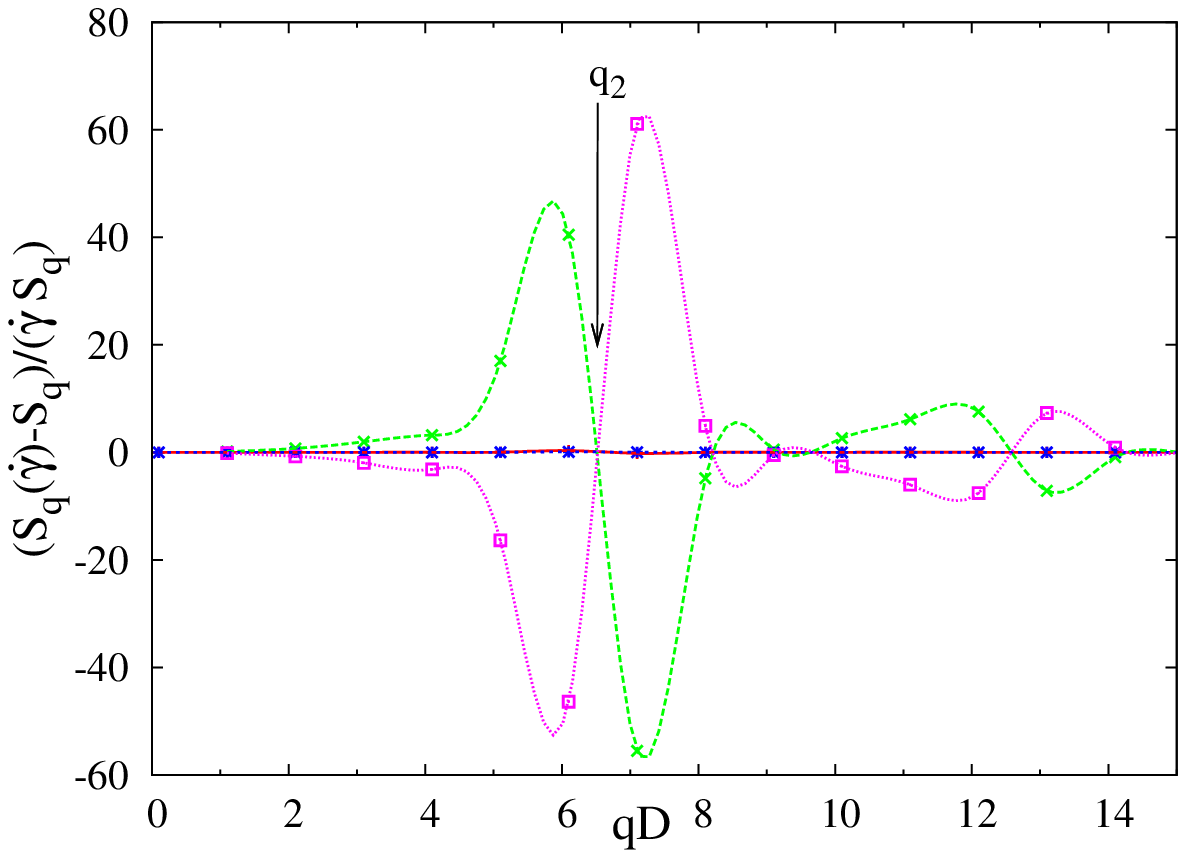}\\
\includegraphics[ width=0.46\textwidth]{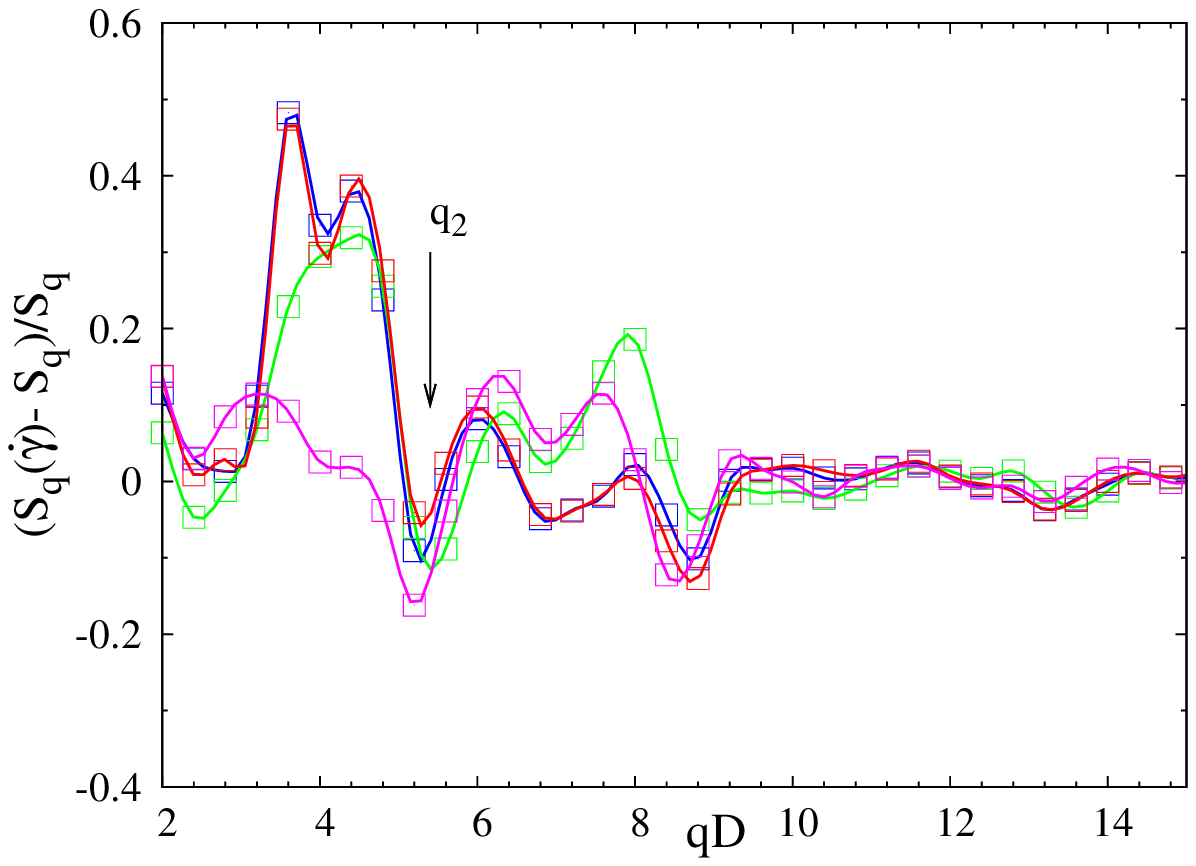}\hfill
\includegraphics[ width=0.46\textwidth]{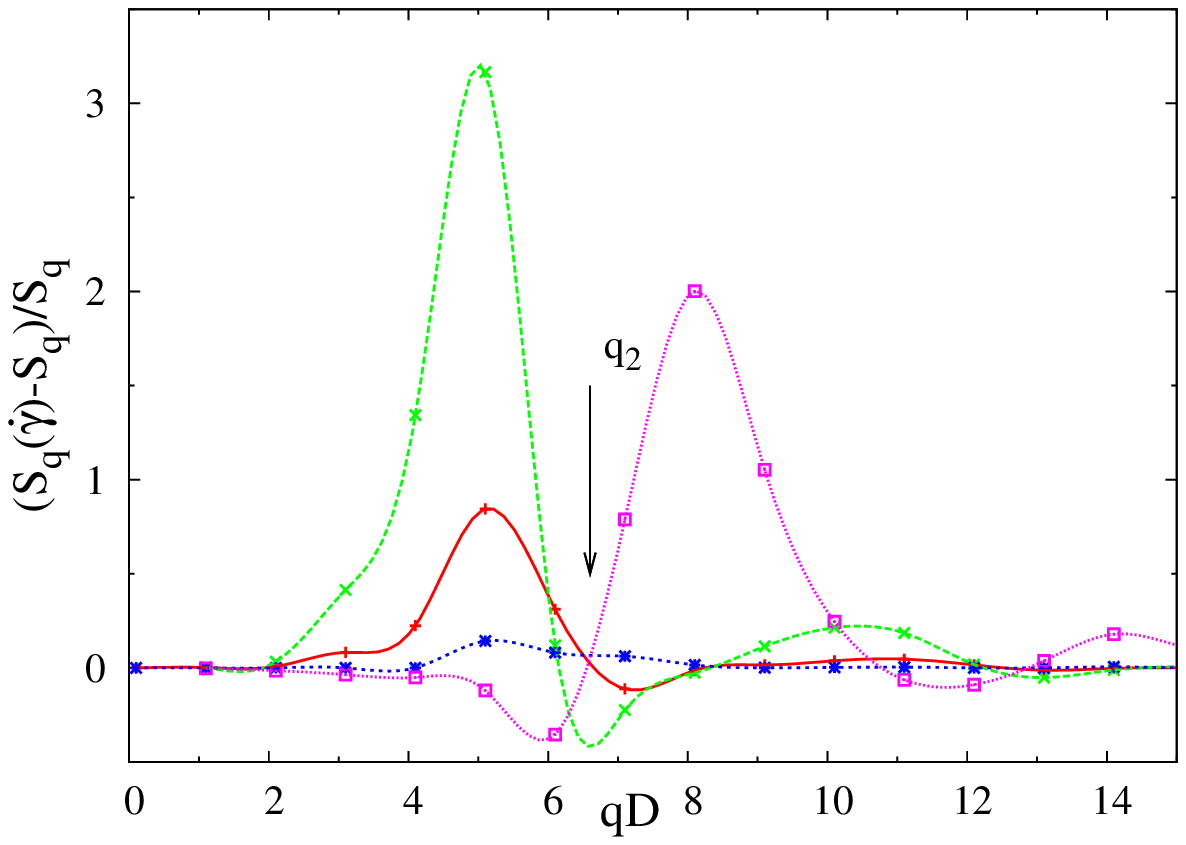}\\
\includegraphics[ width=0.46\textwidth]{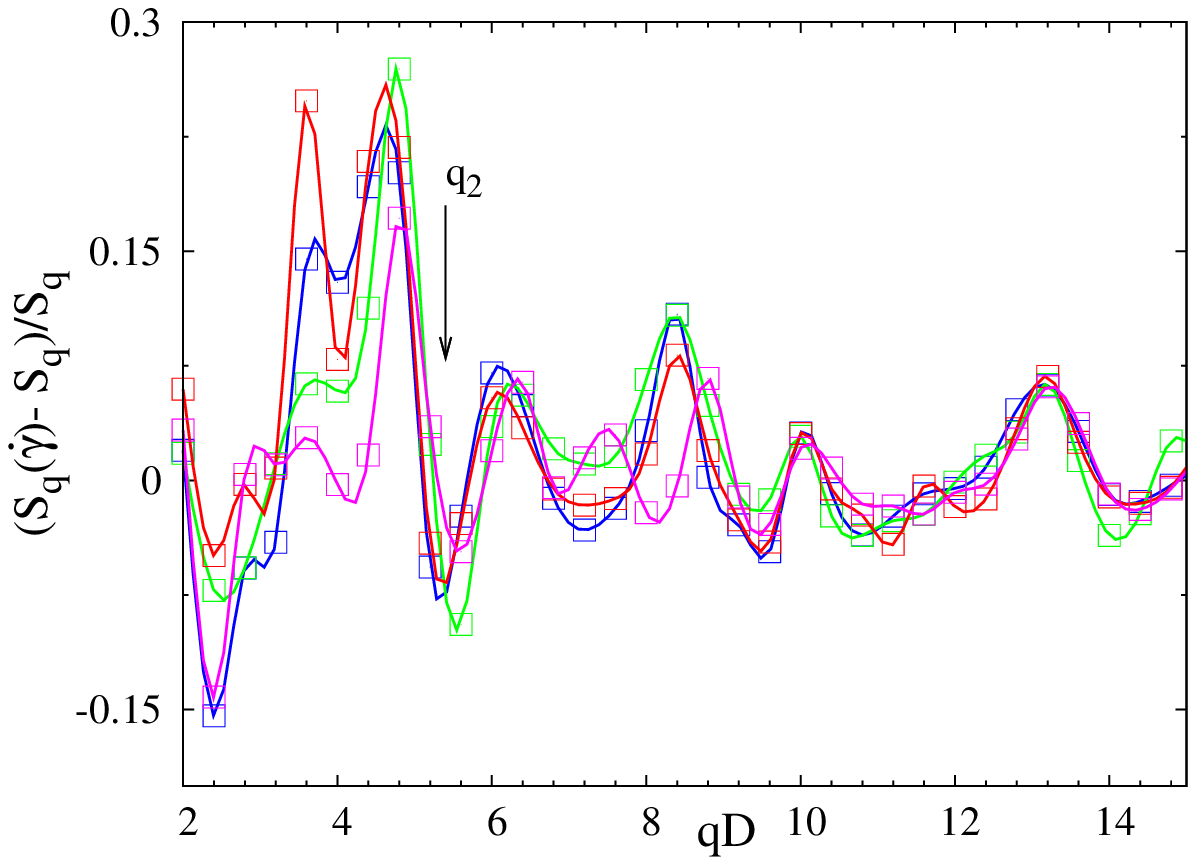}\hfill
\includegraphics[ width=0.46\textwidth]{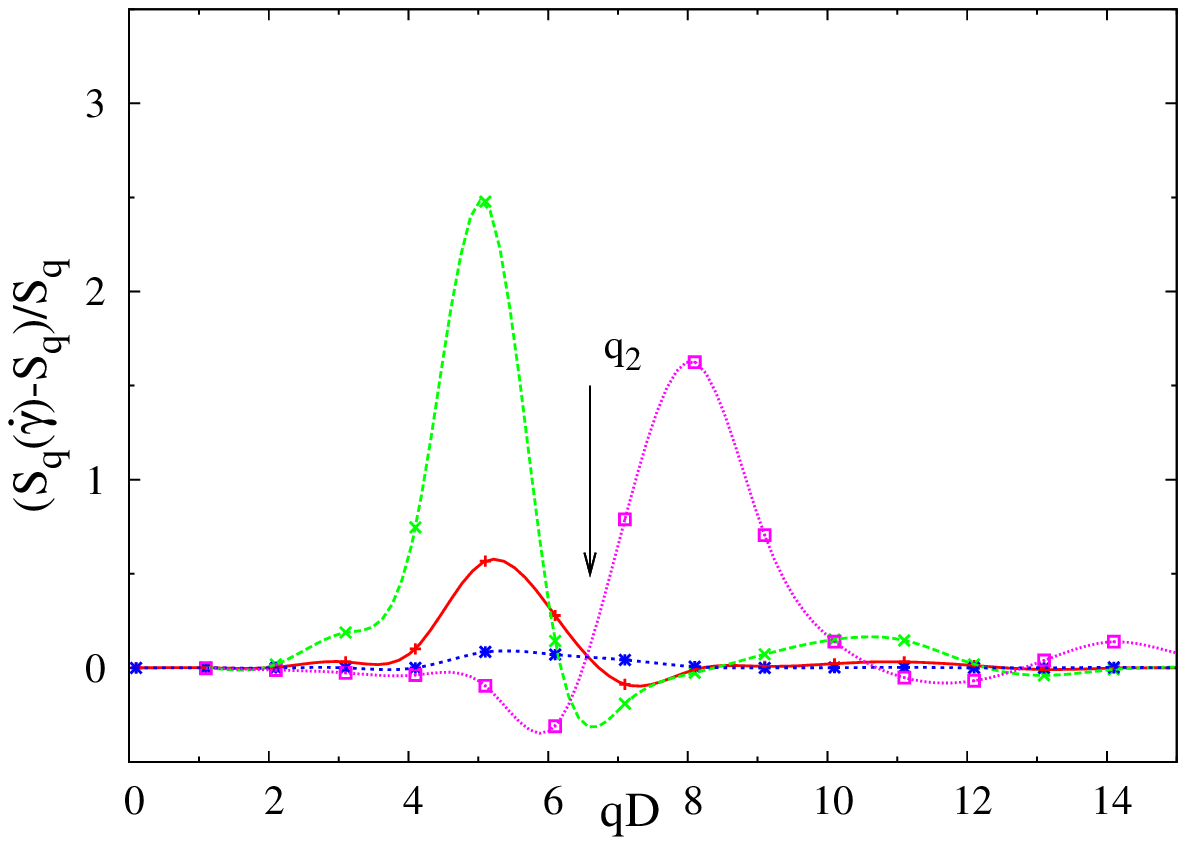}\\
\caption{ Direct comparison of the quantity $(S({\bf
q},\dot{\gamma})-S_q)/S_q$ between simulation and MCT. The selected
orientations are: $q_x=0$ (red), $q_x=q_y$ (green), $q_y=0$ (blue)
and $q_x=-q_y$ (magenta) for $\varphi = 0.79$ in the simulation
(left column). The Pe$_0$ numbers are $4 \cdot 10^{-4}$ (top row), $2 \cdot 10^{-1}$ (middle row) and $2
\cdot 10^{-3}$ (bottom row). For MCT (right column) the values are:
$\epsilon = -10^{-2}$, Pe$_0 = 10^{-4}$ (top row), $\epsilon = 10^{-3}$, Pe$_0 = 10^{-2}$ (middle row) and $\epsilon = 10^{-3}$, Pe$_0 = 10^{-8}$
(bottom row). 
Note that in the linear response case (top row) results were normalized by the shear rate $\dot\gamma$. 
All structure factors are 
averaged over a small angle in the $\bf q$ direction (for an absolute $| {\bm q}|$-value and an angle $\phi$ a small section $\Delta q$ and $\Delta \phi$ is defined; all ${\bm q}$ values lying within the defined section are used for the averaging. $\Delta \phi_{\rm sim} = 5^\circ  $, $\Delta |{\bm q}_{\rm sim}| = 0.4$,$\Delta \phi_{\rm MCT} = 15^\circ  $, $\Delta |{\bm q}_{\rm MCT}| = 1.0$ .).}\label{figsqrel}
\end{figure}

A more careful look onto the distorted microstructure is provided by $q$-dependent cuts through $S_{\bf q}(\dot\gamma)$ along the  directions  colour-coded in Fig. \ref{fig1}. Important for the  direction $q_x=0$ (red)  is the need, present in both simulation and theory, to average $S_{\bf q}(\dot\gamma)$ over a small but finite angle, because exactly at $q_x=0$ the structure oscillates noisily around zero.\\
Especially of interest are the intensities of case  $(iii)$, where the stationary structure of the shear melted solid is studied. The density $\phi=0.79$ is not yet high enough to lie in the glass, but close enough to the glass transition so that the correlations at quite low rate, namely Pe$_0=2\, 10^{-4}$, closely resemble the ones of glassy states at very low shear rates. (While we can get good statistics for stresses at Pe$_0=2\, 10^{-5}$ for all densities, structure factors can not be sampled sufficiently there.) At this density of $\phi=0.79$ also the equilibrium structure factor $S_q$ can be obtained in long simulation runs, and the (relative) difference  $(S_{\bf q}(\dot\gamma,\epsilon)-S_q)/S_q$ can thus be determined. It is shown in Fig. \ref{figsqrel} for two states in order to investigate the distorted structure of the shear melted glass in detail. We base the following discussion on the hypothesis that Pe$_0=2 \cdot 10^{-4}$ at $\phi=0.79$ captures a glass-like state in the limit of low bare Peclet number. 

Figure \ref{figsqrel} provides a sensitive test of the accuracy of the theoretical predictions. The lower left panel shows that the structure factor at vanishing shear rate $S_{\bf q}(\dot\gamma\to0)$ jumps discontinuously at the glass transition; while  $S_{\bf q}(\dot\gamma\to0)\to S_q$ holds in the fluid, $S_{\bf q}(\dot\gamma\to0)\neq S_q$ holds in the glass.   Relative deviations $(S_{\bf q}(\dot\gamma\to0)- S_q)/ S_q$ of 20\% remain. Simulation finds  quite isotropic deviations which show a maximum on the low-$q$ side of the primary peak in $S_q$.  MCT-ITT predicts the absence of a linear response regime in $S_{\bf q}(\dot\gamma)$ as function of the shear rate in the glass, and derives it from the existence of the yield scaling law in the transient correlators. Because $\dot\gamma$ sets the time scale for the final relaxation into the stationary state, the limit $\dot\gamma\to 0$ does not agree with  the quiescent result $\dot\gamma=0$.\\ 
Quantitatively, MCT-ITT overestimates the distortion again by a factor up to 10, and finds a noticeable anisotropy, as discussed in context with Fig. \ref{figsqcolor}. While the difference between the bidisperse and the monodisperse system may influence the comparison, we believe that the major origin of the error is that  MCT-ITT underestimates the speeding up of structural rearrangements caused by shear. The too slow transient correlators thus become anisotropic  because the accumulated strain $\dot\gamma \,t$ becomes too big before structural correlations have decayed. 

Qualitatively, aspects of the anisotropy predicted by MCT-ITT can be seen in the simulations at only slightly larger shear rates, like at Pe$_0= 0.2$ shown in the middle left panel of Fig.~\ref{figsqrel}. While along the two axis- and the extensional diagonal direction, the low-$q$ wing of the primary peak in $S_{\bf q}(\dot\gamma)$ becomes enhanced under shear, along the compressional axis (magenta) it gets suppressed, and the high-$q$ wing is pushed up. Simulation also finds a suppression of the peak height along all directions, which MCT-ITT reproduces along the diagonal directions. Overall the anisotropy and the magnitude of the distortions predicted by MCT-ITT remain too large, but the deviation decreases. The differences between the equilibrium structure factors $S_q$ in the simulated and in the calculated system should be taken into account in future work, and presently preclude comparisons at larger wavevectors.

\section{Conclusions and Outlook}

The first fully quantitative solutions of the MCT-ITT equations for the distorted microstructure and the stresses in steadily sheared two-dimensional shear-thinning fluids and yielding glasses of Brownian hard discs exhibit all the universal features discussed within schematic MCT-ITT models (Fuchs \and Cates 2003). They compare qualitatively well, but quantitatively with appreciable errors, with Brownian dynamics simulations  of a bidisperse mixture without hydrodynamic interactions in a linear shear profile, which for all states considered remains in an homogeneous and disordered state. The non-analytic behavior of the stationary properties, and the lack of a linear response regime throughout the (shear-melted) glass state, predicted by theory, can be found in the simulation.

\section*{Acknowledgements}

Work funded in part by DFG-IRTG 667, and EPSRC grants EP/045316 and EP/030173. MEC holds a Royal Society Research Professorship
\label{lastpage}

\end{document}